\documentclass[twocolumn]{aastex631}

\usepackage{amsmath}

\renewcommand{\vec}[1]{\boldsymbol{#1}}

\newcommand{\twoninethree}{OGLE-2016-BLG-0293}
\newcommand{\sixeightnine}{OGLE-2016-BLG-0689}
\newcommand{\onesixnine}{OGLE-2019-BLG-0169}
\newcommand{\twooneone}{OGLE-2015-BLG-0211}

\newcommand{\sixsixtwo}{OGLE-2018-BLG-0662}
\newcommand{\foureightthree}{OGLE-2018-BLG-0483}
\newcommand{\onefourfive}{OGLE-2015-BLG-0145}
\newcommand{\onefournine}{OGLE-2015-BLG-0149}
\newcommand{\fouronezero}{OGLE-2018-BLG-0410}
\newcommand{\Spitzer}{{\em Spitzer}}
\newcommand{\Gaia}{{\em Gaia}}

\newcommand\KR[1]{\textcolor{black}{#1}}

\begin{document}

%\title{Massive lens population from Spitzer and Gaia}
\title{Analysis of the full Spitzer microlensing sample I: Dark remnant candidates and Gaia predictions}

\author[0000-0002-9326-9329]{Krzysztof A. Rybicki}
\affiliation{Department of Particle Physics and Astrophysics, Weizmann Institute of Science, Rehovot 76100, Israel}
\author{Yossi Shvartzvald}
\affiliation{Department of Particle Physics and Astrophysics, Weizmann Institute of Science, Rehovot 76100, Israel}
\author{Jennifer C. Yee}
\affiliation{Center for Astrophysics $|$ Harvard \& Smithsonian, 60 Garden St, MS-15 Cambridge, MA 02138, USA}
\author{Sebastiano Calchi Novati}
\affiliation{IPAC, Mail Code 100-22, Caltech, 1200 E. California Blvd., Pasadena, CA 91125, USA}
\author{Eran O. Ofek}
\affiliation{Department of Particle Physics and Astrophysics, Weizmann Institute of Science, Rehovot 76100, Israel}
\author{Ian A. Bond}
\affiliation{Institute of Natural and Mathematical Sciences, Massey University, Auckland 0745, New Zealand}
\collaboration{20}{(Leading Authors)}

\author{Charles Beichman}
\affiliation{IPAC, Mail Code 100-22, Caltech, 1200 E. California Blvd., Pasadena, CA 91125, USA}
\author{Geoff Bryden}
\affiliation{IPAC, Mail Code 100-22, Caltech, 1200 E. California Blvd., Pasadena, CA 91125, USA}         
\author{Sean Carey}
\affiliation{IPAC, Mail Code 100-22, Caltech, 1200 E. California Blvd., Pasadena, CA 91125, USA}
\author{Calen Henderson}
\affiliation{IPAC, Mail Code 100-22, Caltech, 1200 E. California Blvd., Pasadena, CA 91125, USA}     
\author{Wei Zhu}
\affiliation{Department of Astronomy, Tsinghua University, Beijing 10084, China}
\author[0000-0002-9113-7162]{Michael~M.~Fausnaugh}
\affiliation{Department of Physics \& Astronomy, Texas Tech University, Lubbock TX, 79410-1051, USA}
\author{Benjamin Wibking}
\affiliation{Michigan State University, 426 Auditorium Road East Lansing, MI 48824, USA}
\collaboration{20}{(The Spitzer Team)}

\author{Andrzej Udalski}
\affiliation{Astronomical Observatory, University of Warsaw, Al. Ujazdowskie 4, 00-478, Warszawa, Poland}
\author{Radek Poleski}
\affiliation{Astronomical Observatory, University of Warsaw, Al. Ujazdowskie 4, 00-478, Warszawa, Poland}
\author{Przemek Mr{\'o}z}
\affiliation{Astronomical Observatory, University of Warsaw, Al. Ujazdowskie 4, 00-478, Warszawa, Poland}
\author{Micha{\l} K. Szyma{\'n}ski}
\affiliation{Astronomical Observatory, University of Warsaw, Al. Ujazdowskie 4, 00-478, Warszawa, Poland}
\author{Igor Soszy{\'n}ski}
\affiliation{Astronomical Observatory, University of Warsaw, Al. Ujazdowskie 4, 00-478, Warszawa, Poland}
\author{Pawe{\l} Pietrukowicz}
\affiliation{Astronomical Observatory, University of Warsaw, Al. Ujazdowskie 4, 00-478, Warszawa, Poland}
\author{Szymon Koz{\l}owski}
\affiliation{Astronomical Observatory, University of Warsaw, Al. Ujazdowskie 4, 00-478, Warszawa, Poland}
\author{Jan Skowron}
\affiliation{Astronomical Observatory, University of Warsaw, Al. Ujazdowskie 4, 00-478, Warszawa, Poland}
\author{Krzysztof Ulaczyk}
\affiliation{Department of Physics, University of Warwick, Coventry CV4 7 AL, UK}
\affiliation{Astronomical Observatory, University of Warsaw, Al. Ujazdowskie 4, 00-478, Warszawa, Poland}

\author{Patryk Iwanek}
\affiliation{Astronomical Observatory, University of Warsaw, Al. Ujazdowskie 4, 00-478, Warszawa, Poland}
\author{Marcin Wrona}
\affiliation{Astronomical Observatory, University of Warsaw, Al. Ujazdowskie 4, 00-478, Warszawa, Poland}
\collaboration{50}{(The OGLE Collaboration)}

\author[0000-0001-9823-2907]{Yoon-Hyun Ryu} 
\affiliation{Korea Astronomy and Space Science Institute, Daejeon 34055, Republic of Korea}

\author[0000-0003-3316-4012]{Michael D. Albrow}
\affiliation{University of Canterbury, School of Physical and Chemical Sciences, Private Bag 4800, Christchurch 8020, New Zealand}

\author[0000-0001-6285-4528]{Sun-Ju Chung}
\affiliation{Korea Astronomy and Space Science Institute, Daejeon 34055, Republic of Korea}

\author{Andrew Gould} % No OrcID on purpose
\affiliation{Max-Planck-Institute for Astronomy, K\"onigstuhl 17, 69117 Heidelberg, Germany}
\affiliation{Department of Astronomy, Ohio State University, 140 W. 18th Ave., Columbus, OH 43210, USA}

\author[0000-0002-2641-9964]{Cheongho Han}
\affiliation{Department of Physics, Chungbuk National University, Cheongju 28644, Republic of Korea}

\author[0000-0002-9241-4117]{Kyu-Ha Hwang}
\affiliation{Korea Astronomy and Space Science Institute, Daejeon 34055, Republic of Korea}

\author[0000-0002-0314-6000]{Youn Kil Jung}
\affiliation{Korea Astronomy and Space Science Institute, Daejeon 34055, Republic of Korea}
\affiliation{National University of Science and Technology (UST), Daejeon 34113, Republic of Korea}

\author[0000-0002-4355-9838]{In-Gu Shin}
\affiliation{Center for Astrophysics $|$ Harvard \& Smithsonian, 60 Garden St.,Cambridge, MA 02138, USA}

\author[0000-0003-0626-8465]{Hongjing Yang}
\affiliation{Department of Astronomy, Tsinghua University, Beijing 100084, China}

\author[0000-0001-6000-3463]{Weicheng Zang}
\affiliation{Center for Astrophysics $|$ Harvard \& Smithsonian, 60 Garden St.,Cambridge, MA 02138, USA}

\author[0000-0002-7511-2950]{Sang-Mok Cha}
\affiliation{Korea Astronomy and Space Science Institute, Daejeon 34055, Republic of Korea}
\affiliation{School of Space Research, Kyung Hee University, Yongin, Kyeonggi 17104, Republic of Korea} 

\author{Dong-Jin Kim}
\affiliation{Korea Astronomy and Space Science Institute, Daejeon 34055, Republic of Korea}

\author{Hyoun-Woo Kim} % ONLY for events with years <= 2019
\affiliation{Korea Astronomy and Space Science Institute, Daejeon 34055, Republic of Korea}

\author[0000-0003-0562-5643]{Seung-Lee Kim}
\affiliation{Korea Astronomy and Space Science Institute, Daejeon 34055, Republic of Korea}

\author[0000-0003-0043-3925]{Chung-Uk Lee}
\affiliation{Korea Astronomy and Space Science Institute, Daejeon 34055, Republic of Korea}

\author[0009-0000-5737-0908]{Dong-Joo Lee}
\affiliation{Korea Astronomy and Space Science Institute, Daejeon 34055, Republic of Korea}

\author[0000-0001-7594-8072]{Yongseok Lee}
\affiliation{Korea Astronomy and Space Science Institute, Daejeon 34055, Republic of Korea}
\affiliation{School of Space Research, Kyung Hee University, Yongin, Kyeonggi 17104, Republic of Korea}

\author[0000-0002-6982-7722]{Byeong-Gon Park}
\affiliation{Korea Astronomy and Space Science Institute, Daejeon 34055, Republic of Korea}

\author[0000-0003-1435-3053]{Richard W. Pogge}
\affiliation{Department of Astronomy, Ohio State University, 140 West 18th Ave., Columbus, OH  43210, USA}
\affiliation{Center for Cosmology and AstroParticle Physics, Ohio State University, 191 West Woodruff Ave., Columbus, OH 43210, USA}

\collaboration{50}{(The KMTNet Collaboration)}

\author{Fumio Abe}
\affiliation{Institute for Space-Earth Environmental Research, Nagoya University, Nagoya 464-8601, Japan}
\author{Richard Barry}
\affiliation{Code 667, NASA Goddard Space Flight Center, Greenbelt, MD 20771, USA}
\author{David P.~Bennett}
\affiliation{Code 667, NASA Goddard Space Flight Center, Greenbelt, MD 20771, USA}
\affiliation{Department of Astronomy, University of Maryland, College Park, MD 20742, USA}
\author{Aparna Bhattacharya}
\affiliation{Code 667, NASA Goddard Space Flight Center, Greenbelt, MD 20771, USA}
\affiliation{Department of Astronomy, University of Maryland, College Park, MD 20742, USA}
\author{Akihiko Fukui}
\affiliation{Department of Earth and Planetary Science, Graduate School of Science, The University of Tokyo, 7-3-1 Hongo, Bunkyo-ku, Tokyo 113-0033, Japan}
\affiliation{Instituto de Astrofisica de Canarias, Via Lactea s/n, E-38205 La Laguna, Tenerife, Spain}
\author{Ryusei Hamada}
\affiliation{Department of Earth and Space Science, Graduate School of Science, Osaka University, Toyonaka, Osaka 560-0043, Japan}
\author{Shunya Hamada}
\affiliation{Department of Earth and Space Science, Graduate School of Science, Osaka University, Toyonaka, Osaka 560-0043, Japan}
\author{Naoto Hamasaki}
\affiliation{Department of Earth and Space Science, Graduate School of Science, Osaka University, Toyonaka, Osaka 560-0043, Japan}
\author{Yuki Hirao}
\affiliation{Institute of Astronomy, Graduate School of Science, The University of Tokyo, 2-21-1 Osawa, Mitaka, Tokyo 181-0015, Japan}
\author{Stela Ishitani Silva}
\affiliation{Department of Physics, The Catholic University of America, Washington, DC 20064, USA}
\affiliation{Code 667, NASA Goddard Space Flight Center, Greenbelt, MD 20771, USA}
\author{Yoshitaka Itow}
\affiliation{Institute for Space-Earth Environmental Research, Nagoya University, Nagoya 464-8601, Japan}
\author{Rintaro Kirikawa}
\affiliation{Department of Earth and Space Science, Graduate School of Science, Osaka University, Toyonaka, Osaka 560-0043, Japan}
\author{Naoki Koshimoto}
\affiliation{Department of Earth and Space Science, Graduate School of Science, Osaka University, Toyonaka, Osaka 560-0043, Japan}
\author{Yutaka Matsubara}
\affiliation{Institute for Space-Earth Environmental Research, Nagoya University, Nagoya 464-8601, Japan}
\author{Shota Miyazaki}
\affiliation{Institute of Space and Astronautical Science, Japan Aerospace Exploration Agency, 3-1-1 Yoshinodai, Chuo, Sagamihara, Kanagawa 252-5210, Japan}
\author{Yasushi Muraki}
\affiliation{Institute for Space-Earth Environmental Research, Nagoya University, Nagoya 464-8601, Japan}
\author{Tutumi NAGAI}
\affiliation{Department of Earth and Space Science, Graduate School of Science, Osaka University, Toyonaka, Osaka 560-0043, Japan}
\author{Kansuke NUNOTA}
\affiliation{Department of Earth and Space Science, Graduate School of Science, Osaka University, Toyonaka, Osaka 560-0043, Japan}
\author{Greg Olmschenk}
\affiliation{Code 667, NASA Goddard Space Flight Center, Greenbelt, MD 20771, USA}
\author{Clement Ranc}
\affiliation{Sorbonne Universit\'e, CNRS, UMR 7095, Institut d'Astrophysique de Paris, 98 bis bd Arago, 75014 Paris, France}
\author{Nicholas J. Rattenbury}
\affiliation{Department of Physics, University of Auckland, Private Bag 92019, Auckland, New Zealand}
\author[0000-0002-1228-4122]{Yuki K. Satoh}
\affiliation{College of Science and Engineering, Kanto Gakuin University, 1-50-1 Mutsuurahigashi, Kanazawa-ku, Yokohama, Kanagawa 236-8501, Japan}
\author{Takahiro Sumi}
\affiliation{Department of Earth and Space Science, Graduate School of Science, Osaka University, Toyonaka, Osaka 560-0043, Japan}
\author{Daisuke Suzuki}
\affiliation{Department of Earth and Space Science, Graduate School of Science, Osaka University, Toyonaka, Osaka 560-0043, Japan}
\author{Paul . J. Tristram}
\affiliation{University of Canterbury Mt.\,John Observatory, P.O. Box 56, Lake Tekapo 8770, New Zealand}
\author{Aikaterini Vandorou}
\affiliation{Code 667, NASA Goddard Space Flight Center, Greenbelt, MD 20771, USA}
\affiliation{Department of Astronomy, University of Maryland, College Park, MD 20742, USA}
\author{Hibiki Yama}
\affiliation{Department of Earth and Space Science, Graduate School of Science, Osaka University, Toyonaka, Osaka 560-0043, Japan}
\collaboration{50}{(MOA collaboration)}

\author{{\L}ukasz Wyrzykowski}
\affiliation{Astronomical Observatory, University of Warsaw, Al. Ujazdowskie 4, 00-478, Warszawa, Poland}

\author{Kornel Howil}
\affiliation{Astronomical Observatory, University of Warsaw, Al. Ujazdowskie 4, 00-478, Warszawa, Poland}

\author{Katarzyna Kruszy{\'n}ska}
\affiliation{Las Cumbres Observatory, 6740 Cortona Drive, Suite 102, Goleta, CA 93117, USA}

\begin{abstract}

In the pursuit of understanding the population of stellar remnants within the Milky Way, we analyze the sample of $\sim 950$ microlensing events observed by the Spitzer Space Telescope between 2014 and 2019. In this study we focus on a sub-sample of nine microlensing events, selected based on their long timescales, small microlensing parallaxes and joint observations by the Gaia mission, to increase the probability that the chosen lenses are massive and the mass is measurable.
Among the selected events we identify lensing black holes and neutron star candidates, with potential confirmation through forthcoming release of the Gaia time-series astrometry in 2026. Utilizing Bayesian analysis and Galactic models, along with the Gaia Data Release 3 proper motion data, four good candidates for dark remnants were identified: OGLE-2016-BLG-0293, OGLE-2018-BLG-0483, OGLE-2018-BLG-0662, and OGLE-2015-BLG-0149, with lens masses of $2.98^{+1.75}_{-1.28}~M_{\odot}$, $4.65^{+3.12}_{-2.08}~M_{\odot}$, $3.15^{+0.66}_{-0.64}~M_{\odot}$ and $1.4^{+0.75}_{-0.55}~M_{\odot}$, respectively. Notably, the first two candidates are expected to exhibit astrometric microlensing signals detectable by Gaia, offering the prospect of validating the lens masses.
The methodologies developed in this work will be applied to the full Spitzer microlensing sample,  populating and analyzing the time-scale ($t_{\rm E}$) vs. parallax ($\pi_{\rm E}$) diagram to derive constraints on the population of lenses in general and massive remnants in particular.

\end{abstract}

\keywords{Microlensing, Stellar remnants, Black Holes, Neutron Stars, Spitzer, Gaia}

\section{Introduction}
\label{sec:intro}

Detecting and characterizing stellar remnants is vital for our understanding of the evolution of stars and populations in the Milky Way. Black holes, in particular, have been of great interest due to their role in the growth and formation of galaxies. Neutron stars offer unique insights into stellar evolution, extreme matter and fundamental physics. In addition, both can play a role in the distribution of dark matter in galaxies, which is still one of the biggest unresolved mysteries in astrophysics.

Detecting stellar remnants poses a challenging problem due to their typically dim and elusive nature. Historically, efforts to identify these remnants have relied heavily on indirect methods, that require the remnant to be in a binary system. For instance, the discovery of binary systems with compact objects, such as X-ray binaries, has provided crucial evidence of their existence. Notably, the detection of gravitational waves, pioneered by the Laser Interferometer Gravitational-Wave Observatory (LIGO) and the Virgo collaboration (e.g.  \citealt{Abbott2016}), marked a monumental advancement in the direct detection of stellar remnants, particularly black holes and neutron stars.
%Although, majority of the gravitational waves events are caused by black holes of far greater masses than what we expect from stars in the Milky Way. It suggests that, if they are indeed of stelalr origin, they might have formed from very low metallicity stars that lose significantly less mass throught their evolution, which allows for massive supernovae progenitors and thus more massive remnants.

Still, single, \KR{non-accreting}, stellar-mass black holes, as well as aged, isolated neutron stars, are practically inaccessible to date. Consequently, our knowledge about these celestial objects remains limited, although they hold valuable information related to stellar formation and evolution. The only practical way to observe them is through microlensing, i.e. by detecting their gravitational influence on the light from another object. Measuring the mass of the lensing object and  constraining the flux that it emits allows us to assess if it could be a stellar remnant. The only known isolated, stellar-mass black hole was identified using this technique (\citealt{Sahu2022}, \citealt{Lam2022}, \citealt{Mroz2022}, \citealt{Lam_Lu2023}).

Microlensing is inherently limited in terms of mass measurement due to its reliance on obtaining the Einstein radius $\theta_{\rm E}$ from
\begin{equation}
    \label{eq:mass}
    M_{\rm L}=\frac{\theta_{\rm E}}{\kappa \pi_{\rm E}},~~\pi_{\rm E}=\frac{\pi_{\rm rel}}{\theta_{\rm E}}, ~~\pi_{\rm rel}=\frac{1}{D_{\rm L}} - \frac{1}{D_{\rm s}}
\end{equation}
where $\kappa=8.144~\mathrm{mas}/M_{\odot}$, $\pi_{\rm E}$ is the magnitude of the microlensing parallax vector (see e.g. \citealt{Gould2004}), $D_{\rm L}$ is the distance to the lens and $D_{\rm s}$ is distance to the source.

Direct measurements of $\theta_{\rm E}$ are challenging to obtain in practice. In some cases it can be done through observations of caustic crossings, resolving the images using interferometry (\citealt{Delplancke2001}, \citealt{Cassan2016}) or astrometric microlensing \citep{Dominik2000}. All these approaches are somewhat sub-optimal for studying large populations of objects. The first requires very special circumstances to occur: a caustic-crossing of either a binary lens, or less likely, the central passage of a single lens in front of the source. The second, resolving the images with high-precision interferometers, is a promising avenue. Two such detections have been reported to date: for the Kojima event (\citealt{Nucita2018}), \citealt{Fukui2019}, \citealt{Zang2020}) and Gaia19bld (\citealt{Cassan2022}, \citealt{Bachelet2022}, \citealt{Rybicki2022}). Although this technique is limited to brighter targets, we should expect an increasing number of interferometric observations of microlensing events thanks to developments in the field (\citealt{Gravity+}, \citealt{Gould2023}).
The third approach to direct $\theta_{\rm E}$ measurement is through astrometric microlensing, which requires a sub-miliarcsecond astrometric precision. This limitation will also slowly be overcome, especially thanks to the development of advanced space satellites like Gaia (e.g. \citealt{Rybicki2018}, \citealt{Kluter2020}) and Roman (e.g. \citealt{Roman_BH_2023}, \citealt{Fardeen2024}). It could also be possible to investigate seeing-limited data sets like OGLE or KMTNet (Segev et al. in prep.), where the number of measurements might help to overcome limited astrometric accuracy. However, up to now, success in observing astrometric microlensing has been limited. There are ongoing attempts to detect it using adaptive optics (e.g. \citealt{Lu2016}), but only limits were obtained from this kind of studies. The signal has been detected only in handful of events using the Hubble Space Telescope and under very special circumstances (\citealt{Sahu2017}, \citealt{Zurlo2018}, \citealt{Sahu2022}, \citealt{McGill2023})\footnote{Gaia preliminary astrometric time series also confirmed the light centroid deviations in the Gaia16aye event \url{https://www.cosmos.esa.int/web/gaia/iow_20210924}}. As of today, it is still a technique that can only be applied to specific cases.

An alternative method for estimating the mass of the lens in microlensing events involves using the timescale of the event and microlensing parallax, while making assumptions about the distribution of proper motions, because $\theta_{\rm E} = \mu_{\rm rel}t_{\rm E}$. Then, one would calculate the mass as
\begin{equation}
    M_{\rm L} = \frac{\mu_{\rm rel}t_{\rm E}}{\kappa\pi_{\rm E}} = 1.35M_{\odot} \left[ \frac{\mu_{\rm rel}}{4\,\rm mas/yr}\right] \left[\frac{t_{\rm E}}{100\,\rm d}\right] \left[\frac{0.1}{\pi_{\rm E}}\right].
    \label{eq:mass2}
\end{equation}
However, this approach is reliant on assumptions about the proper motion distribution and galactic model, which can introduce uncertainties and biases in the final mass estimates. It is important to note, that with this approach one does not directly measure the mass of the lens, and only recovers statistical information about this parameter, depending on the assumed Galactic model. As mentioned before, it is also necessary to constrain the flux coming from the lens to be able to claim that it is a dark stellar remnant.

Several studies have already employed this mass estimation approach, where statistical information about $\mu_{\rm rel}$ is applied. For example, \cite{Wyrzykowski2016} applied this method to the OGLE-III sample of 59 events exhibiting a parallax signal. Later, additional analysis was conducted for the same sample of events, but it also implemented the source proper motion from the Gaia Data Release 3 (GDR3) catalog, to tighten the prior on the relative proper motion distribution \citep{WyrzykowskiMandel2020}. In the following studies, \cite{Mroz+Wyrzykowski2021} refined and improved the technique, while \cite{Mroz2021} applied it again to the OGLE-III data set.

The crucial element of this approach hinges on the detection or, at the very least, constraint of the microlensing parallax signal within the selected sample's events. Unfortunately, the ground-based-only photometric measurements usually do not provide strong constraints on the microlensing parallax, as it requires the Earth accelerated motion around the Sun to be significant.

In addition, the more pronounced (and thus - easier to measure) the microlensing parallax signatures, the higher the value of the $\pi_{\rm E}$ parameter. This bias poses an even greater challenge in the examination of massive stellar remnants, as events with smaller microlensing parallaxes tend to favor more massive lenses (see Equation \ref{eq:mass}). One way to avoid such bias and also measure smaller $\pi_{\rm E}$ values, would be to rely on simultaneous space satellite observations to identify the microlensing parallax signal, instead of Earth's orbital motion.
%The key ingredient of this approach is that the microlensing parallax signal in the events from the chosen sample has to be detected, or at least constrained. Unfortunately, this is not the case for most of the microlensing events that are found in the surveys, which strongly limits the sample. Moreover, those events that allow for good measurement of this parameter, usually exhibit stronger microlensing parallax signatures, meaning that such samples are biased toward events with higher values of $\pi_{\rm E}$. This effect is even more problematic for the studies of massive stellar remnants - as one can immediately see from equation \ref{eq:mass}, events with smaller microlensing parallaxes favor more massive lenses.

To utilize this idea, we reviewed the sample of $\sim 950$ events that have been a part of the Spitzer microlensing campaign, which was conducted in the 2014-2019 seasons. The Spitzer campaign was directed specifically towards the goal of extra-solar planet characterization. Nonetheless, the procedure of target selection and the observing strategy for the campaign is well defined \citep{Yee2015Spitzer}, meaning it is a controlled sample, which allows drawing conclusions about the general stellar remnant population.
In this study, we do not explore the whole population of stellar remnants based on the Spitzer sample, but rather select particular candidates with longer timescales and smaller values of the $\pi_{\rm E}$ parameter, which are likely to be caused by a more massive lens. Then, after assessing the amount of light that is emitted by the blend and the lens, one can judge whether the lens is a good candidate for a dark stellar remnant or not. Furthermore, we select and analyze only those events that were observed by the \Gaia\, mission, as they might be verifiable in the near future, thanks to high precision astrometric measurements from \Gaia\, that could be used to measure or constrain $\theta_{\rm E}$.
%Out of the total sample of $\sim$950 Spitzer events, we selected nine events that are good candidates to host a remnant lens,
%So far there were only studies of particular, interesting cases - mostly planetary. The two more general works have been done on small sub-samples (\citealt{Calchi.2015.B}, \citealt{Zhu2017}).

This paper is organized as follows. In Section \ref{sec:tEpiE} we present an initial review and analysis of the whole Spitzer microlensing sample, which is performed to populate an initial $\lowercase{t}_{\rm E}$ -- $\pi_{\rm E}$ diagram, necessary to select candidates for stellar remnants lenses. In Section \ref{sec:massive_candidates} we do an in-depth analysis of the nine events that were selected, including the derivation of lens mass and distance distributions, which utilizes our light curve analysis and priors on lens-source proper motions based on the Milky Way models. In Section \ref{sec:gaia_predictions} we simulate realistic Gaia astrometry for a range of possible $\theta_{\rm E}$ values, including the most probable ones resulting from the analysis presented in Section \ref{sec:massive_candidates}. We summarize and give conclusions in Section \ref{sec:summary}.

%\section{Spitzer microlensing sample}
\section{Populating the $\lowercase{t}_{\rm E}$ -- $\pi_{\rm E}$ plane}
\label{sec:tEpiE}
To find the sub-sample of microlensing events hosting stellar remnants, we first need to populate the $\lowercase{t}_{\rm E}$ -- $\pi_{\rm E}$ diagram, which will allow us to select potentially massive lenses.
First, as we only consider standard (point source, point lens; hereafter PSPL) events with parallax, we filter out all the events with clear caustic-crossing or finite source signatures \KR{based on visual inspection}. After this step, out of the $\sim 950$ Spitzer events, $720$ remain. These events constitute our final sample that is used for the construction of the $\lowercase{t}_{\rm E}$ -- $\pi_{\rm E}$ diagram.
% Spitzer Calendar
%#Week Targets Due Cutoff Time   Obs Start     Obs End
% 1    7553.15	  7552.50	7557.50	      7564.50
% 2    7560.15	  7559.50	7564.50	      7571.50
% 3    7567.15	  7566.50	7571.50	      7578.50
% 4    7574.15	  7573.50	7578.50	      7585.50
% 5    7581.15	  7580.50	7585.50	      7598.50
\subsection{Ground-based data}

During the modeling procedure (see Section \ref{sec:light_curve_model_1}) we are fitting the PSPL model with parallax to the joint set of OGLE+KMT+\Spitzer\, data. In this step of unsupervised, automatic fitting we decide to omit MOA data for practical reasons, as OGLE+KMT sets are sufficient for creating an initial model. Later on, in a detailed analysis of individual events from the selected sub-sample, we use the full re-reduced data.

Triggering of targets in the Spitzer microlensing campaign was based on the OGLE EWS (\citealt{Udalski1992}, \citealt{Udalski2015}) and MOA \citep{Bond2001} alerts. The KMTNet \citep{Kim2016} data were incorporated into the decision-making starting in 2016.
The OGLE-IV data were collected with a large mosaic camera, consisting of 32 CCD chips with resolution 2048x4096 pixels each, and the scale of 0.26 arcsec/px. The camera is mounted on the 1.3-meter Warsaw Telescope, located in Las Campanas Observatory in Chile. The cadence for each event varied between $\sim 4/\rm hr$ up to $\sim 0.5/\rm day$, depending on the field, with the frequency of observations declining with the increasing (projected) distance from the Galactic center. The data were reduced with the OGLE-IV photometric pipeline \citep{Udalski2015}, which implements an improved DIA procedure from \cite{Wozniak2000}. Full OGLE-IV light curves were used, with the data collected up until\footnote{Which marks the beginning of the 2020 bulge season and is a practical cut-off date for observations of 2019 (and earlier) events, given that OGLE paused its operations because of the COVID-19 outbreak at this time.} $HJD'\approx8920$\footnote{$HJD' \equiv HJD-2450000$}.

The KMTNet survey uses three 1.6-meter telescopes, located in Cerro Tololo Inter-American Observatory (CTIO, Chile), Siding Spring Observatory (SSO, Australia) and South African Astronomical Observatory (SAAO, South Africa). Each telescope has a mosaic camera with 4, 9k x 9k CCD chips mounted and a pixel scale of 0.4 arcsec/px. As the majority of events have good OGLE coverage, in this initial step of building an initial $\lowercase{t}_{\rm E}$ -- $\pi_{\rm E}$ diagram it is sufficient to use publicly available KMTNet data from the automatic pipeline, which is only the part of the light curve from the discovery season for each event. In the detailed analysis of smaller sub-sample we use all the available, re-reduced data.

\subsection{\Spitzer\, data}

%As the main goal of the campaign was planet population in the Milky Way, the planetary candidates were favored and thus the sample is biased, which certainly will complicate the further analysis of the global sample. Nonetheless, the triggering and target selection was well defined and described in \cite{Yee2015Spitzer}.
%Due to some delay between the alert of the potentially interesting target and Spitzer observations, the satellite data usually only cover the slope of the light curve - most of the time the decline, due to the configuration between the Spitzer and Earth location during the time of observation.
The Spitzer Space Telescope microlensing campaign was conducted during the ``warm" part of the mission and so the data were collected using the IR, narrow bandwidth $L$, centered at the wavelength of $3.6~\mu \rm m$.
While covering the peak of the event is the most beneficial for the microlensing parallax determination, the satellite observations usually only cover a part of the light curve, as there is a delay between the target selection and the actual observation time. Naturally, it is also not known \textit{a priori} when exactly the event is going to peak from the Spitzer perspective.
%- most of the time the decline - due to the delay between the target selection and the actual observation time.

Each year, targets could have been observed by Spitzer  during either of the two, $\sim$38 day long windows (Northern summer and winter), when the Galactic bulge was visible from the satellite's location \citep{Calchi.2015.A}. In practice, most of the events were only observed within the summer window, as during the winter the Galactic center is not accessible for Earth-based instruments. There were a few exceptions, where the event was either observed in multiple years (mostly for the baseline information), or during the winter window. In fact, a few events from the selected ``massive" sub-sample did get such additional winter observations, which significantly enhanced the microlensing parallax measurements (see Section \ref{sec:massive_candidates}). Nonetheless, in the full, 950-event sample of Spitzer microlensing events, most of the targets have only one patch of Spitzer data, taken within the summer window, with $\sim1$ day cadence, over a period of $\sim2-5$ weeks.

The rough description and numbers quoted above, while not detailing exact information for every specific event, provide the needed understanding of the Spitzer light curves’ coverage that is sufficient for the goals of our work. The specific details of the Spitzer microlensing campaign, its observing strategy, target selection criteria are beyond the scope of this paper. More information can be found in e.g. \cite{Yee2015Spitzer} and \cite{Udalski2015Spitzer}. The detailed analysis of the whole Spitzer microlensing sample, which will be used for more general studies of the population of the Galactic planets and stellar remnants, will be presented in a separate paper.

\subsection{Light curve modeling}
\label{sec:light_curve_model_1}

To construct the initial $\lowercase{t}_{\rm E}$ -- $\pi_{\rm E}$ plane and identify the events with potentially massive lenses, we fit a standard PSPL microlensing model with parallax to the 720 events that were not classified as ``clearly binary" during the initial, visual inspection. In our light curve analysis we use procedures from the \verb|pyLIMA| package \citep{pyLIMA2017} which employs \verb|VBBinaryLensing| code for light curve computation \citep{VBB2018}.
%events that were not classified as "clearly binary" during the initial, visual inspection. Thus, the following modeling procedure was applied to the sample of $\sim700$ PSPL (Point Source, Point Lens) events. Since in this step we aim to populate a initial $\lowercase{t}_{\rm E}$ -- $\pi_{\rm E}$ space and only filter-out potential candidates for massive stellar remnants, we do not need all the available photometric data. Thus, we decide to not include the MOA data sets in this initial step.

First, we re-scale error bars in all the data sets, which is necessary to obtain meaningful parameter uncertainties and to compare different models. We apply the re-scaling procedure in steps to consecutive groups of data sets, fixing the error bars in the groups that have already been modified. We start with fitting to the OGLE data as it has the most stable photometry and long baseline. We first fit a PSPL model, which is then used as a seed for the two ($u_0>0$ and $u_0<0$) fits of PSPL+parallax models. Out of the these two fits we pick the one with the better $\chi^2$ to be a reference. We then apply a standard formula for the new error bars (e.g. \citealt{Yee2012}):
\begin{equation}
    \sigma_{\rm new} = \sqrt{(\gamma\sigma_{\rm old})^2 + \epsilon^2},
\end{equation}
where we fix the value of the error floor $\epsilon=0.005\, \mathrm{mag}$ for all data sets but \Spitzer, for which we do not set the floor for the error. We find the re-scaling factor $\gamma$ by requiring $\chi^2/dof = 1$. This procedure is repeated multiple times, which iteratively removes outliers. After the whole procedure is finished for the OGLE data, we add all the other ground-based sets, and finally, the \Spitzer\, data.
%then \Gaia\, data and finally \Spitzer\,. In Table \ref{tab:rescaling} we list the re-scaling factors for all the events that were later included in the "massive" sub-sample.

Starting with a simple least-squares minimization, we fit a PSPL model
%with annual parallax
to the re-scaled ground-based data.
%separately exploring $u_0>0$ and $u_0<0$ space.
We calculate the model using the standard formula for the magnification (e.g. \citealt{Paczynski1986}):
\begin{equation}
    A(u)=\frac{u^2+2}{u\sqrt{u^2+4}};~~u(t)=\sqrt{u_0^2+\left(\frac{t-t_0}{t_{\rm E}}\right)^2},
\end{equation}
where $t_0$, $u_0$ and $t_{\rm E}$ are the standard microlensing parameters: time of maximum, smallest projected separation in the units of Einstein radius and Einstein timescale, respectively. The magnification enters the formula for total flux that is changing during the event:
\begin{equation}
    F_{\rm tot} (t) = A(t) F_{\rm s} + F_{\rm bl},
    \label{eq:F_tot}
\end{equation}
where $F_{\rm s}$ is the flux from the source and $F_{\rm bl}$ is flux from the blend. During the modeling procedure we include the blend flux through the blending parameter
\begin{equation}
    g=\frac{F_{\rm bl}}{F_{\rm s}}.
\end{equation}

We use the derived Paczyński parameters ($t_0$, $u_0$, $t_{\rm E}$) as a seed in the next fitting step. Then, before the final step, we fit a model with parallax to the ground-based data only, to assess the microlensing parallax signal resulting from Earth's orbital motion. The model incorporates two additional parameters: North and East components of the microlensing parallax vector $\vec{\pi_{\rm E}}=(\pi_{\rm EN}, \pi_{\rm EE})$ (see e.g. \citealt{Gould2004} for details). Finally, the joint fit to all space and ground-based data is performed, where both annual, and space parallax effects have to be taken into account. 

The space parallax also allows constraints on the $(\pi_{\rm EN}, \pi_{\rm EE})$ vector. Having a satellite at projected distance $D_{\perp}$ from Earth, allows calculation of the microlensing parallax based on the difference between the $t_0$ and $u_0$ parameters measured from the ground and from space (e.g. \citealt{Refsdal1966}, \citealt{Gould1994_space}):
\begin{equation}
    \vec{\pi_{\rm E}} = \frac{\rm au}{D_{\perp}}(\Delta \tau, \Delta u_0),
\end{equation}
where
\begin{equation}
    \Delta \tau=\frac{t_{0,sat}-t_0}{t_{\rm E}},~~\Delta u_0=u_{0,sat}-u_0,
\end{equation}
and the $sat$ subscript refers to parameters measured from the perspective of the satellite. The space-based measurement of the microlensing parallax is independent from the ground-based measurement, which allows for an additional cross-check between the two.

\begin{figure*}
    \centering
    \includegraphics[width=\textwidth]{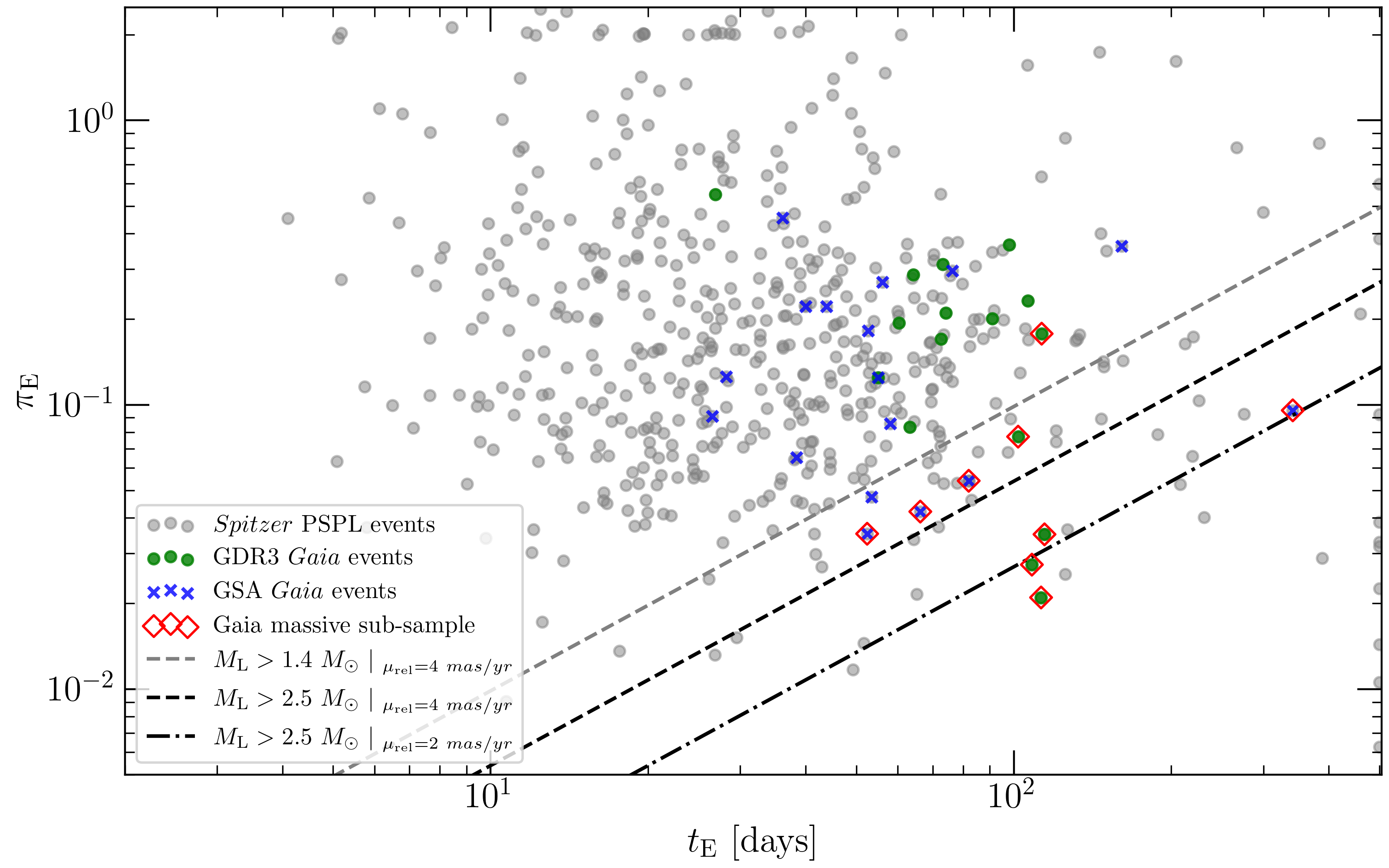}
    \caption{Initial $\lowercase{t}_{\rm E}$ -- $\pi_{\rm E}$ plane, populated with the results from parallax PSPL fit to all Spitzer events (gray points). Colored crosses and circles mark all the events that were also observed by Gaia mission and published either through Gaia Science Alerts (blue) or in \citealt{Wyrzykowski2023} (green). Red diamonds mark those events that were selected for detailed analysis as they host potentially massive lenses. For a given value of relative lens-source proper motion, one can draw a line of constant lens mass. Assuming a typical value of $\mu_{\rm rel} = 4$ mas/yr, all the events lying below the gray line would be caused by lenses heavier than $1.4~M_{\odot}$, while those below the black line are caused by lenses heavier than $2.5~M_{\odot}$. \KR{To visualize the impact of proper motion, we also plot the dash-dotted line for $\mu_{\rm rel} = 2$ mas/yr and $M_{\rm L} = 2.5~M_{\odot}$. We note that there are two groups of events with faulty fits: one with high values of $t_{\rm E}$ ($\sim500$d) and the other with $\pi_{\rm E}$ ($\sim2$) - see Section \ref{sec:tEpiE} for details.}}
    \label{fig:tE-piE}
\end{figure*}

In both ground-based-only and full parallax fitting we explore $u_0>0$ and $u_0<0$ regimes separately, which usually results in two, similarly well fitted models. While in principle we do expect up to four solutions (the so called four-fold degeneracy, see e.g. \citealt{Gould2004}), at this stage of analysis we limit ourselves only to search for the $u_0+ \leftrightarrow u_0-$ degeneracy. Because during the fitting procedure we start from $\vec{\pi_{\rm E}} = (0,0)$, we expect the solution to converge to the smaller values first, and thus we might not be finding solutions with large $\pi_{\rm E}$ values. While for complete analysis of each event the full grid search has to be done, we accept this limitation in this work, as our goal is to find potentially massive lenses, i.e. those with smaller parallaxes. Then, for the selected nine candidates we explore the full $\pi_{\rm EN}-\pi_{\rm EE}$ space (see Section \ref{sec:massive_candidates} \KR{for detailed analysis and selection criteria}).

\subsection{Initial $\lowercase{t}_{\rm E}$ -- $\pi_{\rm E}$ plane}
\label{sec:tEpiE}

Having PSPL models with parallax fitted to all the non-binary light curves from the Spitzer sample of microlensing events, we were able to build an initial $\lowercase{t}_{\rm E}$ -- $\pi_{\rm E}$ plane (Figure \ref{fig:tE-piE}). All the events have two solutions, as for each event the $u_0-$ and $u_0+$ planes were separately explored (see Section \ref{sec:light_curve_model_1}). As in further analysis we are specifically interested in massive lens candidates, in Figure \ref{fig:tE-piE} we are plotting the solution that results in a higher expected mass of the lens (see Equation \ref{eq:mass}).
%With black crosses we mark the events that were not observed by OGLE or KMT, as we require multi-color photometry for the detailed analysis (see Section \ref{sec:massive_candidates}) and they will not be selected for the final sub-sample anyway.
We mark the events that were observed by the Gaia mission (blue crosses and green circles). From among them we choose the ones that host potentially massive lens (red diamonds) - those are analyzed in detailed as good candidates for dark remnants (see Section \ref{sec:massive_candidates}).

The majority of the events lie approximately in the middle of the presented space, but one might identify two distinct, smaller groups: one with large microlensing parallax values, clumped at $\pi_{E}\approx 2$ and one with long timescales of $t_{\rm E} \approx 500$ days, which is the internal, upper limit on this parameter value in the fitting algorithm. Events in both groups (in total $\approx40$) suffer from faulty fits. After visual inspection of the light curves and fitting results, we conclude that the reason is either low SNR of ground-based data, significant systematic trends (likely due to proper motion of the source/lens, which impacts DIA reduction procedure), or a low number of Spitzer data points. All of these factors, and particularly their combination, lead to inaccurate microlensing parallax measurements, especially if the effect is small. As a result, the fitting algorithm converges to parameter values that are inaccurate or wrong, or does not converge at all. While most of these issues can be addressed either by careful re-modeling, de-trending or simply filtering-out impacted light curves, such analysis is beyond the scope of this paper. There are no events observed by \Gaia\, in these groups and so they would not enter the final sample, although we still plot them here for the sake of completeness. 

\section{Massive lens candidates}
\label{sec:massive_candidates}

In this section we describe the analysis of the sub-sample consisting of candidates for potentially massive lenses.

\subsection{Sample selection}

There were three general criteria used during the selection procedure: long Einstein timescale $t_{\rm E}$, low value of microlensing parallax $\pi_{\rm E}$ and presence of Gaia data. The first two together makes higher values of lens mass $M_{\rm L}$ more likely, although we note that it also depends on the lens-source relative proper motion (Equation \ref{eq:mass}). The last condition makes the mass of the lens (potentially) verifiable in the near future, thanks to the Gaia astrometry - see Section \ref{sec:gaia_predictions} for the analysis of this aspect.  
Construction of the initial $\lowercase{t}_{\rm E}$ -- $\pi_{\rm E}$ diagram for all the PSPL Spitzer events allows easy selection of candidates according to the criteria described above. We refrain from applying strict cuts on the parameters, as there is no clear reason for such. Our sub-sample will not be representative for the stellar remnant population and the selection procedure will be arbitrary anyway. We pick mostly events lying below the gray, dashed line on figure {\ref{fig:tE-piE}}, i.e. those for which the more massive solution results in $M_{\rm L}>1.4~M_{\odot}$ (as a lower limit on the mass of neutron stars), under the assumption of $\mu_{\rm rel} = 4~\rm mas/yr$. We supplement this sample with event \onefourfive, which, although it lies above the gray line, has a long timescale of $t_{\rm E} \gtrsim 100~\rm days$ and with its relatively small $\pi_{\rm E}$ value might still be considered as a good candidate.
\begin{table}
  \centering
  \caption{Re-scaling factors $\gamma$ for each telescope, along with the number of photometric data points used.}
  \label{tab:rescaling}

     \begin{tabular}{ccccc}
        \hline
        \hline
        \noalign{\smallskip}
        Event\footnote{\KR{For simplicity, a shortened version of the event names are used in tables and figures throughout the paper: OBXXYYYY$\,\equiv\,$OGLE-20XX-BLG-YYYY.}} &Telescope & $\gamma$ & $N$ points\\
        \noalign{\smallskip}
        \hline
        \noalign{\smallskip}
        OB150145 & OGLE & $1.4$ & $554$ \\
        & Spitzer & $1.6$ & $48$ \\
        & Gaia & $1.7$ & $31$ \\
        \noalign{\smallskip}
        OB150149 & OGLE & $1.4$ & $1653$ \\
        & Spitzer & $4.5$ & $50$ \\
        & Gaia & $5.0$ & $41$ \\
        \noalign{\smallskip}
        OB150211 & OGLE & $1.2$ & $937$ \\
        & Spitzer & $4.7$ & $109$ \\
        & Gaia & $1.5$ & $28$ \\
        \noalign{\smallskip}
        OB160293 & OGLE & $1.6$ & $2718$ \\
        & Spitzer & $3.1$ & $24$ \\
        & Gaia & $1.9$ & $32$ \\
        & KMTS & $1.0$ & $1404$ \\
        & MOA & $2.2$ & $1798$ \\
        \noalign{\smallskip}
        OB160689 & OGLE & $1.4$ & $1162$ \\
        & Spitzer & $2.7$ & $16$ \\
        & Gaia & $1.7$ & $21$ \\
        & KMTA & $1.1$ & $442$ \\
        & KMTC & $1.0$ & $726$ \\
        & KMTS & $1.1$ & $662$ \\
        \noalign{\smallskip}
        OB180410 & OGLE & $1.7$ & $2660$ \\
        & Spitzer & $1.9$ & $29$ \\
        & Gaia & $1.1$ & $80$ \\
        & KMTA & $3.1$ & $190$ \\
        & KMTC & $1.6$ & $369$ \\
        & KMTS & $1.9$ & $179$ \\
        \noalign{\smallskip}
        OB180483 & OGLE & $1.4$ & $1170$ \\
        & Spitzer & $1.9$ & $27$ \\
        & Gaia & $0.6$ & $39$ \\
        \noalign{\smallskip}
        OB180662 & OGLE & $2.3$ & $952$ \\
        & Spitzer & $2.4$ & $44$ \\
        & Gaia & $0.5$ & $59$ \\
        & KMTA & $1.4$ & $395$ \\
        & KMTC & $1.1$ & $821$ \\
        & KMTS & $1.3$ & $359$ \\
        \noalign{\smallskip}
        OB190169 & OGLE & $5.0$ & $955$ \\
        & Spitzer & $2.5$ & $33$ \\
        & Gaia & $1.6$ & $52$ \\
        & KMTS & $2.1$ & $3610$ \\
        \noalign{\smallskip}

        \hline
        \hline
     \end{tabular}
\end{table}

This results in a sub-sample of 9 candidates (red diamonds on Figure \ref{fig:tE-piE}) for stellar remnant lenses, for which we perform a more detailed analysis. With the additional source color analysis described below, we can measure the parallax signal more accurately, which in turn leads to more reliable predictions of the physical parameters of the lens, presented in Section \ref{sec:physical}.

\subsection{Gaia photometric data}
\label{sec:gaia_data}

%All the events in the sub-sample have, by definition, \Spitzer\, and \Gaia\, data.
Currently there are two groups of microlensing events observed by \Gaia\, that have photometric time-series publicly available. The first group consists of all events published through Gaia Science Alerts (\citealt{Hodgkin2013, Hodgkin2021}, henceforth GSA). Once an alert is announced through this channel, all the \Gaia\, photometry for the target is published. The second group is the \Gaia\, DR3 microlensing catalog \citep{Wyrzykowski2023}. Because it was constructed using DR3 data, it only contains measurements collected up to $\sim$ May 2017. In the last column of Table \ref{tab:Gaia_parameters} we mark which events from our sample were alerted by GSA, and thus have a full \Gaia\, light curve available.

Most of the photometric observations from the \Gaia\, satellite are taken in $G$-band, and only these are used in this work during the construction of the light curve model. The photometric measurements for the events detected through GSA do not have uncertainties reported, and so we estimate them based on the Gaia DR2 photometric content and validation paper \citep{Evans2018}.

On average, \Gaia\, comes back to the same field every $\sim 30$ days, and every visit it takes two measurements separated by $\sim 6$ hours (each for one of the mirrors), although this number can vary depending on the gradient of the scanning angle (\citealt{Gaia}). Although such frequency of observations is insufficient to properly cover a microlensing light curve, \Gaia\, photometry is only playing a supplementary role in the process of light curve characterization. In particular, we do not detect any meaningful signal of space parallax in the \Gaia\, data (which is expected as \Gaia\, is in orbit around Lagrange point $L_2$, much closer to the Earth than the \Spitzer\, satellite). On the other hand, the astrometric data it will provide, not only will be crucial for the light centroid shift detection (see Section \ref{sec:gaia_predictions}), but also do not require as high cadence as photometry to be useful, due to much longer effective time scale of the astrometric microlensing signal compared to  its photometric counterpart (e.g. \citealt{Dominik2000}).

\subsection{Additional ground-based data}

In the analysis of the \twoninethree\, event we also included data from MOA collaboration \citep{Bond2001}. The photometric measurements were taken with the 1.8 m telescope located at at Mt. John, New Zealand, using their standard broad $R$ filter.

To determine the color of the source (see Section \ref{sec:Spitzer_color_const}) for the \twooneone\, event, we used the $H$-band measurements from the \textit{ANDICAM} instrument \citep{DePoy2003}, mounted on the 1.3m SMARTS telescope located in the CTIO observatory in Chile. 

\subsection{Source stars}
\begin{figure*}
    \centering
    \includegraphics[width=\textwidth]{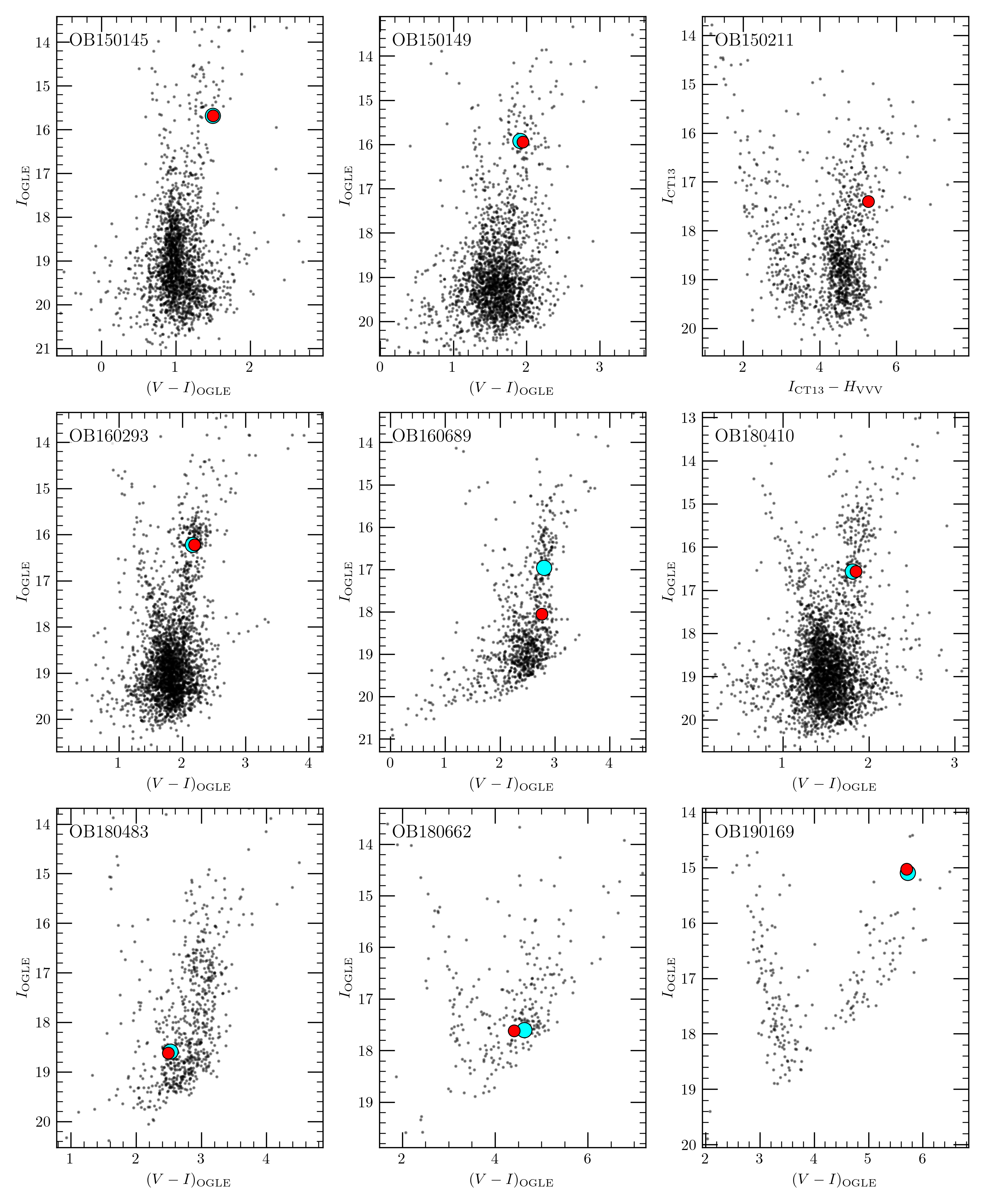}
    \caption{Color-magnitude diagrams for all the massive-lens candidate events, constructed from the indicated catalog of stars (mostly OGLE, in all events but \twooneone) in a $2'\times2'$ box around the magnified object. The large, red dots mark positions of the source stars while the cyan one mark the color and brightness of the baseline (i.e. including blend). They overlap whenever the blending is low, which is the case for almost all events in the sub-sample, excluding \sixeightnine.}
    \label{fig:multipanel_cmd}
\end{figure*}
Generally the information about the color of the source and the position of the source and the blend on the CMD is useful to investigate potential inconsistencies in the model. It is also necessary to support the space parallax calculations based on the \Spitzer\, data (see the next sub-section). To perform a source color analysis, observations of the event in a second band are required. Having regular coverage in two bands that span a reasonably large range of magnifications throughout the event, we are able to derive the source color independently of the model. \KR{Linear regression is used to fit the relationship between the total fluxes in both bands, as the slope of the linear function is equivalent to the source flux ratio in the two bands, which directly translates to the source color} (see Equation \ref{eq:F_tot}). Whenever it is possible, the color analysis is made based on the OGLE data set as it is the one with longest baseline and most reliable photometry. Excluding \twooneone\, all the events in our sample have the OGLE $V$-band available and so the $(V-I)_{OGLE}$ color of the source is derived. Most of the sources reside in the red clump (see Figure \ref{fig:multipanel_cmd}), which suggests that they are red giants that belong to the galactic bulge population. For the case of \twooneone\, we use the follow-up data in $H$-band to construct the CMD diagram and conclude that source is also part of the red clump.
%The light curve model yields the source and blend fluxes. According to the best solution for \twoninethree\, $I_{OGLE, \rm src} =16.21$ mag (see Table \ref{tab:fit_parameters}). Having OGLE $V$-band measurements, we construct $F_V(F_I)$ relation. We follow the standard procedure and using linear regression on said relation, we derive the source color $(V-I)_{\rm src}=2.19 \pm 0.01$. We locate the source in the lower part of the red clump in the color-magnitude diagram, populated with the stars from OGLE catalogue in a $2'\times2'$ arcmin box around the event (see Figure \ref{fig:OB160293_cmd}). It means that the source is most likely a giant star residing in the galactic bulge.
%The baseline color $(V-I)_{\rm base}=2.16\pm0.02$ and $I_{\rm base}=16.22\pm0.01$, which is very close to the source position on the CMD. Indeed, the blending derived from the model is consistent with zero, so we do expect the source and the baseline to be at the same position on CMD. Nonetheless, the exact blend flux is difficult to assess, as the blending parameter $g=F_s/F_{\rm bl}$ is negative. We postpone a more detailed analysis of the blend light to Section \ref{sec:OB160293}.
\subsection{Spitzer color constraints}
\label{sec:Spitzer_color_const}
The \Spitzer\, data often do not allow measurements of the $L$-band baseline flux, which makes the $\pi_{\rm E}$ analysis less constrained.
It is possible to enhance the microlensing parallax information inferred from \Spitzer\, observations by finding the color of the source $(I-L)_{\rm src}$ and using it as a constraint on the Spitzer source brightness $L$ during the modeling process\footnote{$L$ denotes Spitzer's 3.6$\mu$m band.}. To do that, we match stars detected in both OGLE and Spitzer frames and construct a color-color diagram, using $I$, $V$ and $L$ bands (so called $VIL$ relation, see \citealt{Calchi.2015.A} for details). For this purpose we use stars from the red clump, and so we expect such color-color relation to be linear for most of the sources. Using linear regression, we can find its functional form and, by interpolating (or extrapolating) to the known values of the source color $(V-I)_{\rm src}$, derive $(I-L)_{\rm src}$.

Such a constraint is expected to have the highest impact on the light curve fitting for the cases where Spitzer data only constrain the local slope of the light curve (which is often the case). We perform the modeling with and without the constraint and note significant improvement introduced by the color constraint for almost all events. The exception is the case of \sixsixtwo, where Spitzer covered part of the peak, already constraining the shape of the light curve sufficiently.

\subsection{Light curves and modeling}
%% -> grid search for all solutions
%% -> MCMC to find error bars
%% KR: for this event we have only one site data reanalysis from KMT
% figures/tables:
%% Table with final fit parameters (+/- u0 solutions)
%% Light curve figure (KR: Do we want to add panel for residuals from non-parallax model?)
%% piEN piEE grid figure w/wo color constraint (4 panels)
\begin{figure*}
    \centering
    \includegraphics[width=\textwidth]{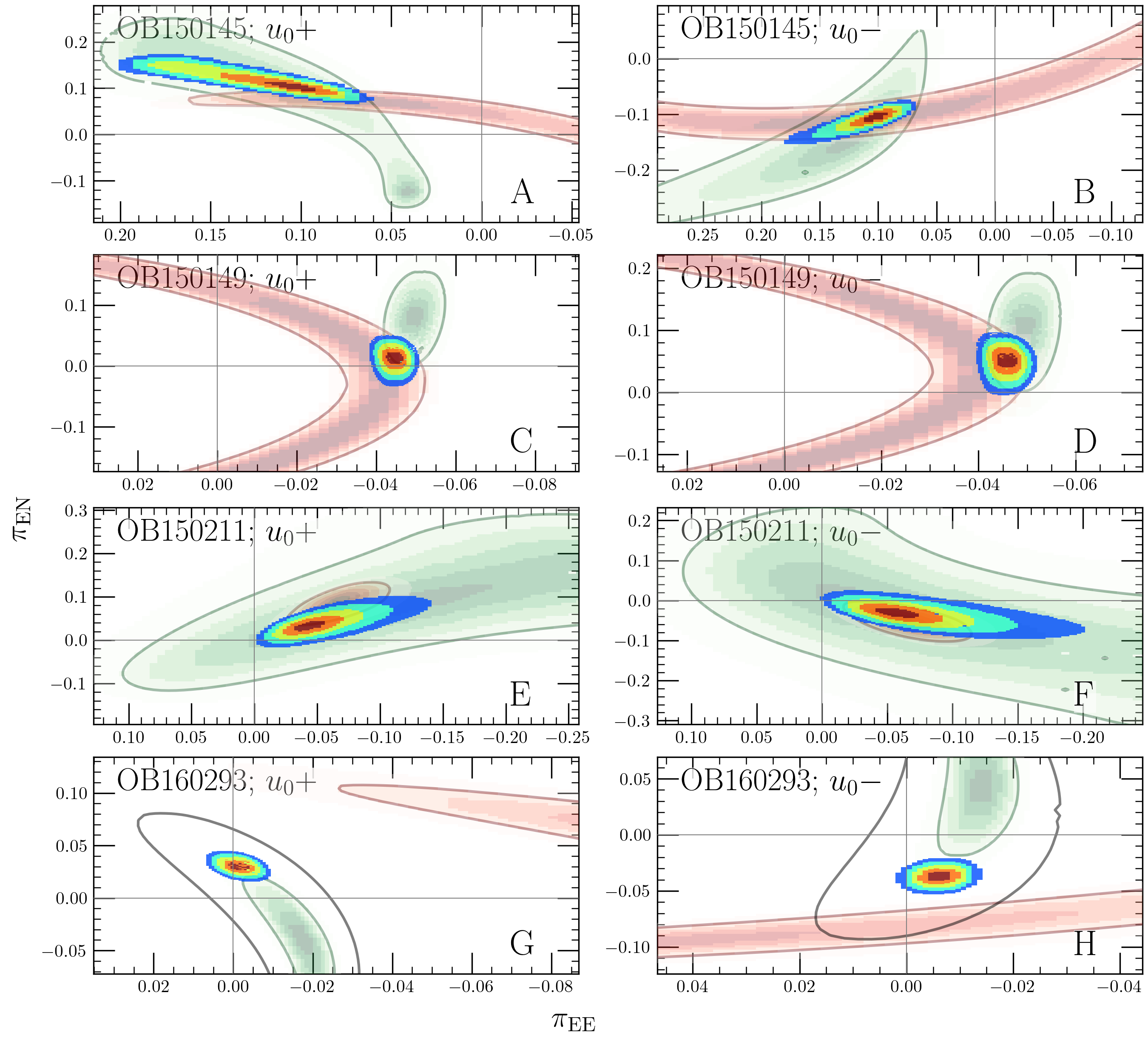}
    \caption{Microlensing parallax contours for $u_0+$ and $u_0-$ solutions for the first four events in the selected sub-sample. Green and red outlines represent the 5--$\sigma$ level for ground-based and Spitzer-``only" (with color constraint) fits, respectively. The colored, filled contours changing from dark red to blue represent 1--5$\sigma$ levels of the final, joint fit. For the events where additional ground-based data sets beyond OGLE were used, we also provide ground parallax contour based only on the OGLE data (gray), to track  potential systematics.}
    \label{fig:multipanel_grids_A}
\end{figure*}
\begin{figure*}
    \centering
    \includegraphics[width=\textwidth]{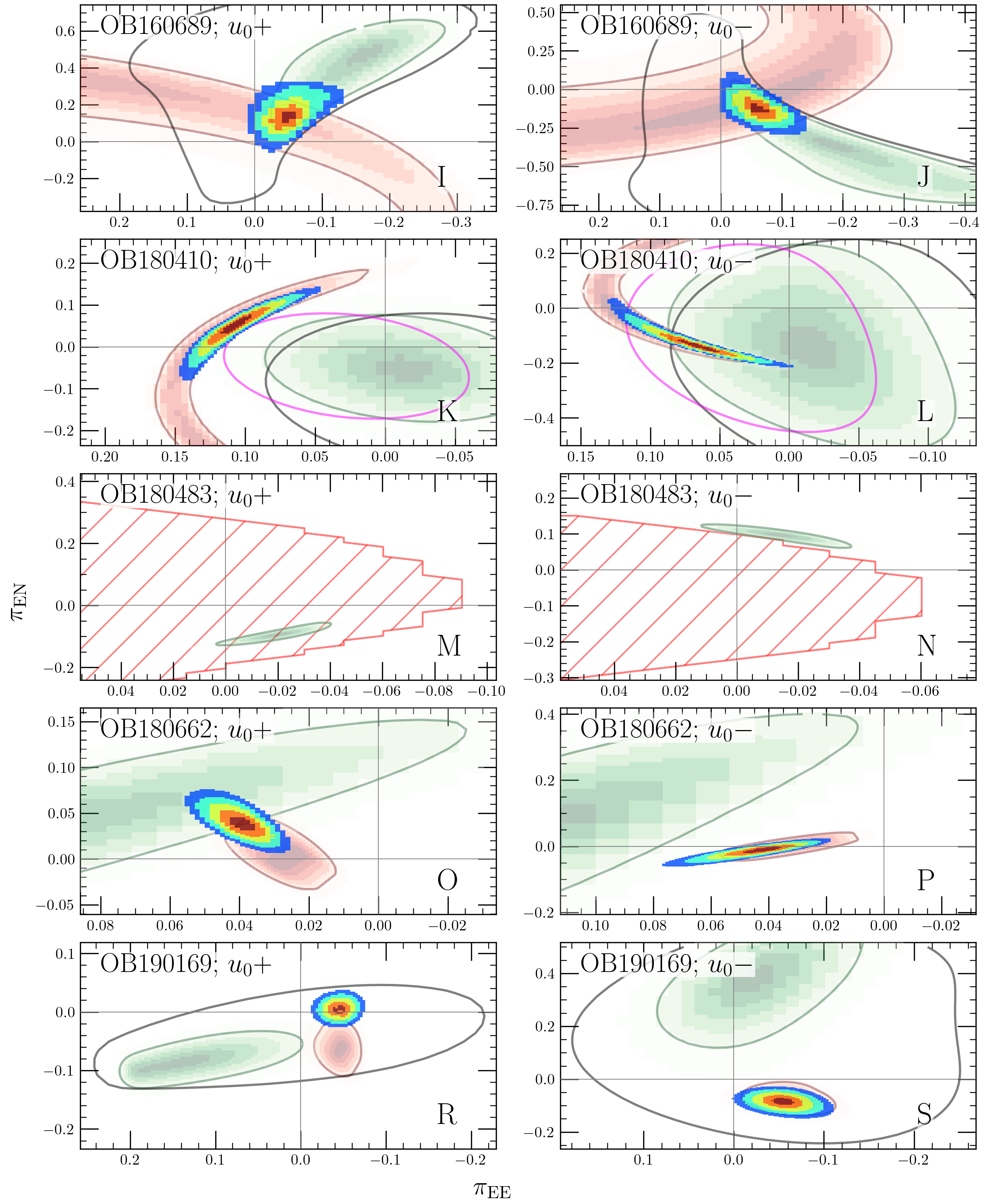}
    \caption{Same as Figure \ref{fig:multipanel_grids_A}, for the remaining 5 events. For the case of \foureightthree\, the \Spitzer\, data allowed only for estimation of the upper limit on the flux, which excludes part of the $\pi_{\rm EN} - \pi_{\rm EE}$ space (marked with red, hatched area on panels $M$ and $N$). The magenta contours on panels $K$ and $L$ represent ground-based-only fits with fixed blending parameter $g=0$.}
    \label{fig:multipanel_grids_B}
\end{figure*}
\begin{figure*}
    \centering
    \includegraphics[width=\textwidth]{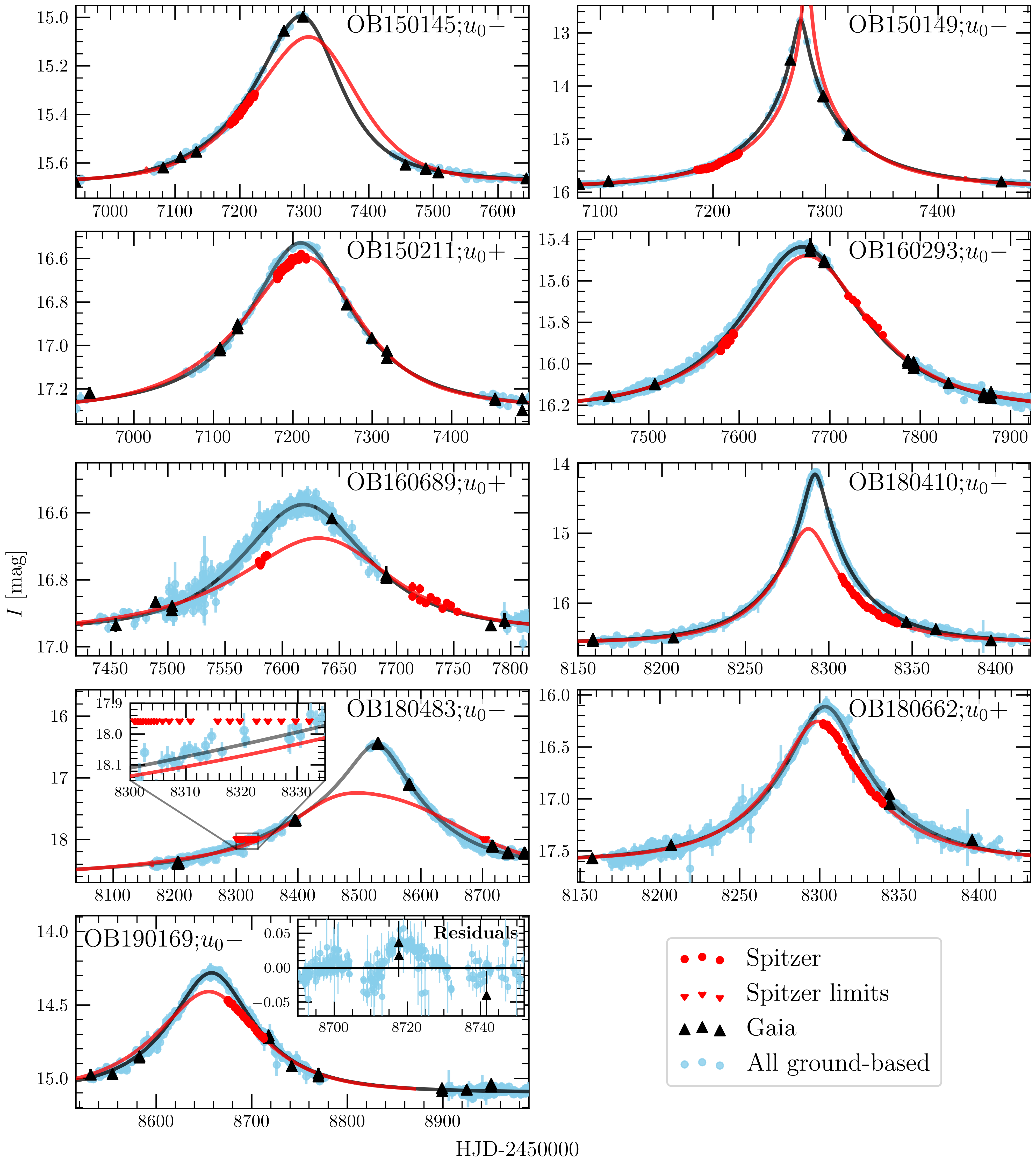}
    \caption{Light curves of all the events from our Spitzer-Gaia sub-sample, with all the data sets used during the modeling. Solid lines represent the best fit of PSPL+parallax model, showing Earth/Gaia (black line) and Spitzer (red line) perspectives. The ground-based data from all the surveys are plotted together (blue points). All models and data points are displayed with respect to $I_{OGLE}$ baseline brightness. For the case of \foureightthree\, we only derive limits on the \Spitzer\, flux which is represented by downward triangles. \KR{Inset on the bottom left panel shows residuals of the region with a small deviation that might have been caused by a low-mass companion to the lens (see details in the text)}.}
    \label{fig:multipanel_lc}
\end{figure*}
 To make sure there are no other solutions due to the four-fold degeneracy \citep{Gould2004}, and to correctly probe the $(\pi_{\rm EN}, \pi_{\rm EE})$ space, we perform a dense grid search over these two parameters. We do the search separately for three fits: ground-based-only, \Spitzer\,-``only" \citep{Gould2020} and joint (using all data sets). The \Spitzer\,-``only" approach is fitting the model to the ground-based data first and then, after fixing $t_0$, $u_0$ and $t_{\rm E}$, fitting $(\pi_{\rm EN}, \pi_{\rm EE})$ and source/blend fluxes using only \Spitzer\, data. Although the accuracy of this approach can be limited, as it does not involve simultaneous fitting of all the parameters, it is useful to gain a better insight into the constraints on the parallax introduced by the ground and space-based data.
In theory, given constraint on the source flux, each \Spitzer\, measurement provides a circular constraint on the $(\pi_{\rm EN}, \pi_{\rm EE})$ plane \citep{Gould-osculating-intersecting}. In practice, the data are taken in different epochs and have a non-zero dispersion. As a consequence, the \Spitzer\,-``only" parallax contours for real events form elongated arcs, which might be reduced further and provide constraints in both dimensions, for the cases where the \Spitzer\, measurements cover larger parts of the light curve or are closer to the peak.

In the Figures \ref{fig:multipanel_grids_A} and \ref{fig:multipanel_grids_B} we gathered relevant parts of the parallax contours for all three modeling approaches. Light curves for the most preferred solutions and a table with the fit parameters are presented in Figure \ref{fig:multipanel_lc} and Table \ref{tab:fit_parameters}, respectively. We present a more detailed discussion and comment on the analysis of each event separately in the following sub-sections.

%Below we provide information about the details of the photometry and light curve modeling for each event individually.
\subsection*{\onefourfive}
\label{sec:OB150145_lc}
%Only OGLE data are available for this event. Spitzer measurements cover the rising slope of the light curve.
The $u_0+$ solution shows nonphysical negative blending level with $g_{OGLE}=-0.7\pm0.12$. The $u_0-$ solution also yields negative blending, but consistent with zero at $g_{OGLE}=-0.06\pm0.12$. The $u_0-$ model is also preferred in terms of the goodness of fit\footnote{The photometry error bars are re-normalized, and so $\chi^2/dof \approx 1$ for all the events (see Table \ref{tab:fit_parameters}). Thus, when comparing different models, we use $\Delta\chi^2$ without quoting the number of $dof$ throughout the text.}, as $\Delta\chi^2\approx18$. In addition, the Spitzer-``only" parallax measurement is slightly more compatible with the ground-based-only measurement for the $u_0-$ case, although for the $u_0+$ solution the two fits also remain in reasonably good agreement.

\subsection*{\onefournine}
While OGLE covered the full event, KMT only observed this event starting in 2016, so there is only the tail of the declining part of the light curve available from this survey.
%The only ground-based data set we use for the analysis of this event is OGLE.
The event lasts for $\sim 100$ days and Spitzer measured only a part of the (rising) slope so most of the parallax information comes from the ground-based light curve, even after including the $VIL$ color constraint. As one can see from panels $C$ and $D$ of Figure \ref{fig:multipanel_grids_A}, Spitzer-``only" and ground contours are consistent.
Blending behaves well and is consistent with zero for both $u_0+$ and $u_0-$ solutions. The $u_0-$ is somewhat preferred according to the photometric models, with $\Delta \chi^2\approx 19$.

\subsection*{\twooneone}
%Only OGLE data are available for this event.
The event lies in a highly extincted field and so there are no measurements in the $V_{OGLE}$ band that could be used to estimate the source color (from GDR3 we can have a very rough estimate of the $V-I$ color as $BP-RP=4.52$ mag). Instead, we used additional observations in $H$-band. The data points were distributed over a large part of the light curve, which allowed for the source color determination.

\Spitzer\, data with a color constraint allows refinement of the microlensing parallax constraints obtained based on the ground-based data (panels $E$ and $F$ in Figure \ref{fig:multipanel_grids_A}). Both $u_0+$ and $u_0-$ solutions have similar $\chi^2$ ($\Delta \chi^2 \approx 2$) and so one can not assess which one is preferred based only on the photometric data.

\subsection*{\twoninethree}
\label{sec:OB160293_lc}
%The \twoninethree\, event was not alerted by Gaia, but it was published in GDR3 microlensing catalog as GaiaDR3-ULENS-088 \citep{Wyrzykowski2023}. As a result, the Gaia photometry is only available until $HJD' \approx 7878$, which is at the end of the decline of the event, less than 0.1 mag above the baseline.

In the OGLE-$I$ and MOA data, short- and long time-scale systematics are visible. That being said, the source is relatively bright, and so some level of baseline variability is expected. In the OGLE data it is comparable to the scatter, and does not seem to have a large impact on the overall fit quality, but in the MOA data, the long time-scale baseline systematics are much more prominent. We address that issue by checking for any color changes in the OGLE $I$-band and $V$-band data, and potential correlations between the baseline color and magnitude changes, which would imply astrophysical origin. The additional analysis do not show such color changes or correlations. Also, trends visible in the MOA data are not compatible with the smaller trends in the OGLE data, and so we conclude that larger systematics visible in MOA should not be taken into account. Thus, we discard the part of the MOA data affected the most and only include the measurements collected at $HJD' > 7000$.

For the case of \twoninethree\, we are able to discriminate between the two solutions, as there are multiple lines of evidence against the $u_0+$ one. First of all, there is a small difference $\Delta \chi^2 \approx 6$, which favors the negative solution. In addition, there is an evident disagreement between the ground-based-only and \Spitzer\,-``only" parallax fits for the $u_0+$ case. Finally, the blending value obtained in the $u_0+$ solution is negative, \KR{and not consistent with zero (4--$\sigma$ away, see Table \ref{tab:fit_parameters})}.

Including additional photometry beyond OGLE data set causes increasing inconsistency between the Spitzer-``only" parallax contours and the ground-based-only contours (see panel $H$ of Figure \ref{fig:multipanel_grids_A}). However, as we mentioned before, the Spitzer-``only" fit is only a diagnostic tool, and a general agreement between the two contours is sufficient to claim that the two sources of the microlensing parallax measurement are consistent for the $u_0-$ case.
%It is worth mentioning that for the $u_0+$ solution, the \Spitzer\,-``only" approach shows another minimum at the North-East quadrant of the $\pi_{\rm EN}$-$\pi_{\rm EE}$ plane, but it is completely excluded by the ground-based data. Overall, the combined data sets provide enough constrains to eliminate any additional solutions, and the only remaining degeneracy is $u_0\leftrightarrow -u_0$, which was known from the initial fits.

\subsection*{\sixeightnine}
\label{sec:OB160689_lc}
%The OGLE and KMTNet survey data are available for this event.
For this event, in addition to the ``regular" Spitzer observations, the data were also taken in the December window , which resulted in covering both the rising and declining part of the light curve. Thanks to that it was possible to determine the microlensing parallax relatively well.

While the blending for both of the solutions is consistent with zero, the posterior distribution for this parameter is also very wide, meaning that the blending parameter is not well constrained from the light curve. Indeed, the blending seems to be present as the source color and baseline color are different (see Figure \ref{fig:multipanel_cmd}). Additionally, there are only 2 points in the OGLE $V$-band on the magnified part of the light curve, which makes the source color determination more uncertain. Thus, we treat the Spitzer source color constraint with caution and conservatively use a wider prior on the Spitzer source color during the fitting (3--$\sigma$ instead of 1--$\sigma$ resulting from the linear regression procedure described in Section \ref{sec:Spitzer_color_const}).
The $u_0+$ and $u_0-$ solutions are similar in terms of the goodness of fit ($\Delta \chi^2 \approx 4$). The $u_0+$ has significant negative blending, but consistent with zero at $g_{OGLE}=-0.31\pm0.21$.

\subsection*{\fouronezero}
\label{sec:OB180410_lc}
%This event has OGLE, MOA and KMTNet data sets available.
After investigation of the photometric data we noted that MOA data shows large deviations that are not present in the remaining two surveys (especially at the wings of the event), and so we decide not to use it in the analysis.
Spitzer data only cover part of the decline of the light curve and by itself provide rather sparse limits on the microlensing parallax measurements. Nonetheless, performing Spitzer-``only" analysis and including the $VIL$ color constraint, shows that space-based parallax gives two, relatively well defined $\chi^2$ minima in $\pi_{\rm EN}-\pi_{\rm EE}$ space for each of the $u_0+$ and $u_0-$ solutions.
The $u_0-$ is preferred with $\Delta\chi^2\approx 19$, and is also much more convincing in terms of consistency between the Spitzer-``only'' and ground-based-only parallax measurement, which can be seen from panels $K$ and $L$, Figure \ref{fig:multipanel_grids_B}. In addition, the blending is negative in both ground-based-only solutions, \KR{and not consistent with zero (3--4\,$\sigma$ away)}. To address this problem we decide to redo the fitting procedure with blending fixed to zero. Although the fit is clearly worse in terms of goodness (with the $\chi^2$ difference of 8 and 20, see Table \ref{tab:fit_parameters}), the tension between the ground-based-only and Spitzer-``only" parallax disappears, at least for the $u_0-$ solution (see magenta contours on panels $K$ and $L$).

\subsection*{\foureightthree}
%Only OGLE survey data are available for this event.
%The event is somewhat faint with the baseline $I_{OGLE}\approx18.5$.
The event was observed by \Spitzer\, in two seasons, separated by about a year, but \Spitzer\, gives no significant constraints on the parallax. Due to the low signal to noise ratio and crowding of the sky region around the event, we were only able to set limits on the \Spitzer\, flux (see the zoom-in in Figure \ref{fig:multipanel_lc} ). As a result, it was only possible to exclude part of the central region in the $\pi_{\rm EN}-\pi_{\rm EE}$ space (see Figure \ref{fig:multipanel_grids_B}, panels $M$ and $N$).  Still, the microlensing parallax can be accurately determined from the ground, which is not surprising as the event is extremely long - Einstein timescale is either $\sim$275 or $\sim$330 days, depending on the solution. The $u_0+$ solution has negative blending only marginally consistent with zero ($g_{OGLE} = -0.15\pm0.05$), which might suggest that the $u_0-$ solution is the real one. More importantly, the $u_0+$ seems to be excluded by the \Spitzer\, limits (Figure \ref{fig:multipanel_grids_B}, panels $M$ and $N$).

\subsection*{\sixsixtwo}
\label{sec:OB180662_lc}
%This event has been observed by OGLE and KMTNet surveys.
%The CMD is quite extreme as the field is highly extincted. The extinction value of $A_I = 4.48$ from Schlafly was used in dark lens analysis.
Before fitting the final light curve model, the OGLE data were corrected for a small linear trend ($\approx 0.04$\,mag over 10 years), likely caused by the change of the source position compared to the position on the reference image. The correction had a minor effect on the final results of the fit.

The microlensing parallax for the \sixsixtwo\, event is weakly constrained from the ground, but with the addition of Spitzer data, a high precision $\vec{\pi_{\rm E}}$ measurement was achieved, as the satellite covered part of the peak.
In both the $u_0+$ and $u_0-$ solutions, the blending is consistent with zero with small error bars, as in both cases $g=-0.01\pm0.02$, which makes this event a good candidate to host a dark lens (see Section \ref{sec:physical}). Comparison of the $\chi^2$ suggests that the $u_0+$ solution is preferred, as $\Delta \chi^2 \approx 28$. In addition, the Spitzer-``only" and ground-based-only fits are more compatible for that case (see panels $O$ and $P$ of the Figure \ref{fig:multipanel_grids_B}).

\subsection*{\onesixnine}
%This event was observed by both OGLE and KMTNet.
%Similarly to \sixsixtwo\, and \twooneone\, this event lies in a highly extincted field.
The Spitzer measurements were only taken at the decline, but close to the peak, which helped to constrain the microlensing parallax from space.
They played a significant role in the microlensing parallax determination - see panels $R$ and $S$, Figure \ref{fig:multipanel_grids_B}. There is a clear offset between the Spitzer-``only'' and all ground-based parallax contours, while the OGLE-only solution is consistent with with Spitzer-``only''. There is a low-level (amplitude of $\sim 20$\,mmag), irregular variability visible in the light curve, which might be the reason for systematic errors in ground-based microlensing parallax measurements. Nonetheless, the final result obtained from the joint fit is driven mostly by the \Spitzer\, data, and so is not strongly affected by the systematics in the ground-based data.

In the light curve of \onesixnine,\, at $HJD'\approx 8720$, there is a clear deviation from the light curve model lasting $\sim 10$ days \KR{(see inset of the bottom left panel on Figure \ref{fig:multipanel_lc})}. It is not an instrumental effect as it appears in both OGLE and KMTNet data. While the scenario of the deviation being a planetary anomaly should not be completely excluded, the amplitude of the variability mentioned in the previous paragraph is comparable to the ``anomaly". Thus, we conduct the analysis using the single lens model and attribute the feature to the variability of the source.

%While this variability ``contaminates" the microlensing signal, in principle it shouldn't impact the PSPL model a lot, as it is short scale (of the order of ~10 days) and very low amplitude ($<20$ mmag, with the exception of the ``anomaly" which is $\sim 40$ mmag above the PSPL model). 
%In addition, the error bars re-scaling procedure somewhat smears out the variability
%Even though the ground-based microlensing parallax measurement could be disturbed by subtle systematics in the data, in this case the parallax is constrained mostly by Spitzer (see panels $G$ and $H$ in Figure \ref{fig:multipanel_grids}) and so the low level variability should not impact the final result significantly.

We note that one of the Gaia epochs was taken during the ``anomaly". In an unlikely scenario of the deviation being due to a planet, it might be an interesting point of the analysis of the Gaia astrometric data. The planetary scenario will be investigated elsewhere as it is beyond the scope of this paper.

\begin{table*}
  \centering
  \caption{The final (joint ground+space+source color constraint) light curve model fit results for all the events. Most of them have two ($u_0+$ and $u_0-$) solutions with comparable $\chi^2$ but for some it was possible to discriminate between them - see more details in the text. \textbf{Note.} Blending parameter $g$ is the ratio of the blend flux $F_{\rm bl}$ to the source flux $F_{\rm s}$. The baseline brightness $I_0$ is recovered with 1--2 mmag precision in all cases. \KR{The photometry error bars are re-normalized, and so $\chi^2/dof \approx 1$ for all the events. Thus, when comparing different models, we use $\Delta\chi^2$ without quoting the number of $dof$ throughout the text.}}
  \label{tab:fit_parameters}
     \begin{tabular}{ccccccccc}
        \hline
        \hline
        \noalign{\smallskip}
        Event & $t_0-2450000$ & $u_0$ & $t_{\rm E}$ & $\pi_{\rm EN}$ & $\pi_{\rm EE}$ & $I_{0,OGLE}$ & $g_{OGLE}$ & $\chi^2/dof$\\
        &[days]&&[days]&&&[mag]&&\\
        \noalign{\smallskip}
        \hline
        \noalign{\smallskip}
        OB150145&$7297.8^{+1.2}_{-1.1}$&$1.13^{+0.26}_{-0.17}$&$76.6^{+8.0}_{-9.5}$&$0.108^{+0.014}_{-0.010}$&$0.111^{+0.020}_{-0.014}$&15.683&$-0.70^{+0.13}_{-0.12}$&668/622\\
        &$7296.9^{+1.1}_{-1.0}$&$-0.62^{+0.05}_{-0.06}$&$116.9^{+5.2}_{-5.5}$&$-0.106^{+0.007}_{-0.007}$&$0.104^{+0.010}_{-0.009}$&15.684&$-0.06^{+0.13}_{-0.13}$&650/622\\
        \noalign{\smallskip}
        OB150149&$7277.9^{+0.01}_{-0.01}$&$0.05^{+0.0003}_{-0.0003}$&$101.8^{+0.4}_{-0.4}$&$0.013^{+0.008}_{-0.008}$&$-0.045^{+0.001}_{-0.001}$&15.922&$0.02^{+0.01}_{-0.01}$&1790/1733\\
        &$7277.9^{+0.01}_{-0.01}$&$-0.05^{+0.0003}_{-0.0003}$&$100.8^{+0.4}_{-0.4}$&$0.051^{+0.010}_{-0.010}$&$-0.046^{+0.001}_{-0.001}$&15.922&$0.01^{+0.01}_{-0.01}$&1771/1733\\
        \noalign{\smallskip}
        OB150211&$7209.9^{+0.4}_{-0.4}$&$0.65^{+0.09}_{-0.07}$&$95.8^{+7.6}_{-7.9}$&$0.035^{+0.011}_{-0.011}$&$-0.046^{+0.011}_{-0.013}$&17.291&$-0.24^{+0.16}_{-0.15}$&1076/1063\\
        &$7209.5^{+0.4}_{-0.4}$&$-0.64^{+0.07}_{-0.09}$&$96.0^{+7.9}_{-8.0}$&$-0.032^{+0.011}_{-0.012}$&$-0.060^{+0.015}_{-0.018}$&17.291&$-0.21^{+0.16}_{-0.16}$&1078/1063\\
        \noalign{\smallskip}
        OB160293&$7669.7^{+0.2}_{-0.2}$&$0.58^{+0.01}_{-0.01}$&$107.3^{+1.2}_{-1.2}$&$0.031^{+0.003}_{-0.003}$&$-0.001^{+0.001}_{-0.001}$&16.223&$-0.11^{+0.03}_{-0.03}$&6096/5959\\
        &$7669.6^{+0.2}_{-0.2}$&$-0.54^{+0.01}_{-0.01}$&$114.3^{+1.4}_{-1.4}$&$-0.037^{+0.003}_{-0.003}$&$-0.006^{+0.001}_{-0.001}$&16.223&$-0.00^{+0.03}_{-0.03}$&6090/5959\\
        \noalign{\smallskip}
        OB160689&$7618.7^{+0.3}_{-0.3}$&$1.08^{+0.19}_{-0.13}$&$66.0^{+5.5}_{-6.5}$&$0.142^{+0.034}_{-0.033}$&$-0.048^{+0.011}_{-0.012}$&16.954&$-0.31^{+0.22}_{-0.21}$&3059/3012\\
        &$7618.5^{+0.3}_{-0.3}$&$-0.87^{+0.08}_{-0.10}$&$77.5^{+4.7}_{-4.9}$&$-0.131^{+0.033}_{-0.030}$&$-0.063^{+0.014}_{-0.014}$&16.954&$0.09^{+0.20}_{-0.20}$&3063/3012\\
        \noalign{\smallskip}
        OB180410&$8291.8^{+0.01}_{-0.01}$&$0.11^{+0.001}_{-0.001}$&$53.5^{+0.3}_{-0.3}$&$0.052^{+0.021}_{-0.022}$&$0.107^{+0.010}_{-0.011}$&16.567&$-0.02^{+0.01}_{-0.01}$&3599/3490\\
        &$8291.8^{+0.01}_{-0.01}$&$-0.11^{+0.001}_{-0.001}$&$53.0^{+0.3}_{-0.3}$&$-0.136^{+0.022}_{-0.019}$&$0.067^{+0.013}_{-0.014}$&16.567&$-0.03^{+0.01}_{-0.01}$&3580/3490\\
        &$8291.8^{+0.01}_{-0.01}$&$0.11^{+0.0001}_{-0.0001}$&$54.2^{+0.03}_{-0.03}$&$0.015^{+0.019}_{-0.020}$&$0.122^{+0.007}_{-0.007}$&16.567&$-0.00^{+0.00}_{-0.00}$&3607/3490\\
        &$8291.8^{+0.01}_{-0.01}$&$-0.11^{+0.0001}_{-0.0001}$&$54.2^{+0.03}_{-0.03}$&$-0.064^{+0.029}_{-0.023}$&$0.103^{+0.010}_{-0.010}$&16.567&$-0.00^{+0.00}_{-0.00}$&3600/3490\\
        \noalign{\smallskip}
        OB180483&$8532.6^{+0.5}_{-0.5}$&$0.16^{+0.01}_{-0.01}$&$272.6^{+10}_{-10}$&$-0.095^{+0.007}_{-0.007}$&$-0.020^{+0.004}_{-0.004}$&18.584&$-0.15^{+0.05}_{-0.05}$&1170/1225\\
        &$8533.3^{+0.6}_{-0.5}$&$-0.13^{+0.01}_{-0.01}$&$328.1^{+11}_{-10}$&$0.097^{+0.006}_{-0.006}$&$-0.014^{+0.005}_{-0.005}$&18.586&$0.02^{+0.05}_{-0.05}$&1170/1225\\
        \noalign{\smallskip}
        OB180662&$8303.5^{+0.02}_{-0.02}$&$0.26^{+0.004}_{-0.004}$&$65.0^{+0.9}_{-0.8}$&$0.039^{+0.007}_{-0.007}$&$0.040^{+0.003}_{-0.003}$&17.600&$-0.01^{+0.02}_{-0.02}$&2711/2613\\
        &$8303.4^{+0.0}_{-0.0}$&$-0.26^{+0.004}_{-0.004}$&$65.2^{+0.9}_{-0.9}$&$-0.012^{+0.008}_{-0.008}$&$0.043^{+0.006}_{-0.005}$&17.600&$-0.01^{+0.02}_{-0.02}$&2740/2613\\
        \noalign{\smallskip}
        OB190169&$8657.9^{+0.1}_{-0.1}$&$0.48^{+0.01}_{-0.01}$&$79.3^{+1.5}_{-1.5}$&$0.005^{+0.006}_{-0.006}$&$-0.045^{+0.006}_{-0.006}$&15.096&$0.13^{+0.05}_{-0.05}$&4799/4637\\
        &$8658.0^{+0.1}_{-0.1}$&$-0.48^{+0.02}_{-0.02}$&$78.9^{+2.0}_{-1.9}$&$-0.085^{+0.012}_{-0.012}$&$-0.054^{+0.011}_{-0.011}$&15.096&$0.13^{+0.06}_{-0.06}$&4768/4637\\
        \noalign{\smallskip}
        \noalign{\smallskip}
        \hline
        \hline
     \end{tabular}
\end{table*}
\subsection{Physical parameters - methodology}
\label{sec:physical}

To assess the probability that the lens is dark, and to provide reasonable predictions of astrometric signal expected from the Gaia mission (see Section \ref{sec:gaia_predictions}), we need to estimate the physical properties of the lens, namely its mass, distance and brightness. All the events in the sub-sample analyzed here are standard events with microlensing parallax signal, which means that the light curve does not contain enough information to directly measure these properties. Nonetheless, the microlensing model provides some constraints which, coupled with the assumptions about the Milky Way kinematics and structure, can be used to evaluate Bayesian probabilities on the physical properties of the lens.

\subsection*{Mass and distance}
\label{sec:mass_and_distance}
We adopt a similar approach to the one presented in \cite{Kruszynska2022} and \cite{Howil2024}, which in turn is based on the procedure used by \cite{Wyrzykowski2016} and later refined by \cite{Mroz+Wyrzykowski2021}. \KR{The technical details regarding the Milky Way model are summarized in \cite{Mroz+Wyrzykowski2021} and in the Appendix A of \cite{Howil2024}.
Below we provide a qualitative description of the analysis and comment on some aspects relevant for our use case.}

\KR{
The starting point of the procedure are posterior distributions of the light curve parameters, obtained from the MCMC modeling. To calculate the mass of the lens, the photometric model has to be supplemented with the source-lens relative proper motion $\mu_{\rm rel}=\left|\vec{\mu_{\rm rel}}\right|$. Initially it is drawn from a wide, flat distribution $[0,30]\, \rm mas/yr$ and later it is weighted according to the galactic model. Similarly, distance to the source $D_{\rm s}$ is drawn from a flat distribution $[0,15]\, \rm kpc$. Then, for each link of the MCMC chains resulting from the light curve modeling, we can calculate mass of the lens (see Equation \ref{eq:mass2}) and also its distance, as
\begin{equation}
    D_{\rm L} = \left(\theta_{\rm E}\pi_{\rm E} + \frac{1}{D_{\rm s}} \right)^{-1}, \,\,\,\theta_{\rm E}=\mu_{\rm rel}t_{\rm E}
\end{equation}
We then apply the "galactic prior" by weighing the resulting mass and distance using weights of the form \citep{Batista2011}:
\begin{equation}
w_{\rm Gal} = \frac{4}{\rm au}\frac{D_{\rm L}^4\mu_{\rm rel}^4 t_{\rm E}}{\pi_{\rm E}}\nu_{\rm d}f_{\mu}f_M M_{\rm L}.
\end{equation}
The above expression combine three priors: the mass function $f_M$, the relative proper motion prior $f_{\mu}$ and stellar density distribution $\nu_{\rm d}$. Remaining quantities result from the transition between the physical parameters and the microlensing variables.
%(see \citealt{Batista2011} for more details).
The stellar density distribution $\nu_{\rm d}$ consists of two separate expressions, with "double exponential" disk and barred bulge profiles (see \citealt{Batista2011} and \citealt{HanGould2003}).  After applying the weights $w_{\rm Gal}$ to lens mass and distance, we obtain the posterior distributions presented in Figure \ref{fig:DLC_results}.}

\KR{
For the mass function we assume a power law $f_M \sim M^{\alpha}$, and for each event we address the impact of the assumed mass prior on the final distribution by comparing a ``flat" prior $\alpha=-1$ with the Kroupa mass function \citep{Kroupa2001}, where the slope is $\alpha=-2.35$ for the more massive ($M>0.5~M_{\odot}$) tail of the distribution (see Table \ref{tab:DLC_results}). While this is a simplification, as we disregard the different slopes for masses in the range $M_{\rm L} < 0.5~M_{\odot}$, this approach is compatible with the selection process of our sub-sample, which favors more massive lenses. Also, it is a conservative assumption in the context of claims of whether a lens is a dark remnant or not, because the slope of the power law at the massive end is much steeper.}

\KR{The relative proper motion prior $f_{\mu}$ is constructed based on the lens and source proper motion assumptions. For the lens we consider two distinct cases: the lens lying in the galactic disk and in the bulge. In the first scenario we assume a normal distribution $V_{l} \approx N(220, 30)\, \rm km/s$ and $V_{b} \approx N(0, 20)\, \rm km/s$. For the bulge we assume $V_{l}=V_{b} \approx N(0, 100)\, \rm km/s$ (\citealt{HanGould1995}, \citealt{Batista2011}). These distributions are corrected for the motion of the Sun with respect to the local standard of rest \citep{Schonrich2010}. The mean of the disk velocity distribution also can vary as it depends on the distance to the lens. A more detailed description is outlined in \cite{Howil2024}. The physical velocities of a lens are then transformed to proper motions with $\mu_{\rm L}=4.74 V_{\rm L}/D_{\rm L}$ and after subtraction of the source proper motion (see next paragraph) can be used as a final prior $f_{\mu}$.}

The relative proper motion partially depends on the source proper motion, which is available from Gaia DR3 catalog \citep{GaiaDR3}, and we incorporate this information in the prior. Values of GDR3 proper motions for each event are presented in Table \ref{tab:Gaia_parameters}. We also quote the $RUWE$\footnote{Re-normalized Unit Weight Error, see e.g. \cite{GaiaEDR3}. Generally, for a well behaved model $RUWE \approx 1$. As a rule of thumb, $RUWE\gtrsim 1.4$ suggests that the astrometric model might not be reliable.\label{foot:RUWE}} parameter to quantify the credibility of the Gaia measurements. It is important to note that we can use quoted proper motions only for the events where all the light (or at least most of it) actually comes form the source. Otherwise, the proper motion measured by Gaia is a combination of that of the source and blend(s). The majority of events in our sub-sample do not exhibit large blending, and so we decide to include Gaia proper motions in the prior
%as the proper motion of the source whenever the blending parameter $g<0.2$. It is true for the eight out of nine events.
for eight out of nine events.
In the remaining case of \sixeightnine, where the blending appears to be more significant, we use a different approach - see Section \ref{sec:OB160689_MLDL} for more details.

It is also worth noting that the direction of the relative lens-source proper motion can be constrained from the light curve model thanks to the microlensing parallax measurements:
\begin{equation}
    \hat{\mu}_{\rm rel} = \frac{\vec{\mu_{\rm rel}}}{\mu_{\rm rel}} = \frac{\vec{\pi_{\rm E}}}{\pi_{\rm E}},
    \label{eq:piE_direction}
\end{equation}
which has to be taken into account.

%Since the Einstein timescale and microlensing parallax can be derived from the light curve model, the only missing component for the Einstein radius and the lens mass determination is the lens-source relative proper motion $\mu_{\rm rel}$ (see Equation \ref{eq:mass2}).
%Its value depends on the stellar populations (either bulge or disk) that the source and lens belong to, and their respective distance and velocity dispersion. We extract the value of this parameter from the galactic prior. Then, knowing $\theta_{\rm E} = \mu_{\rm rel} t_{\rm E}$, one can derive distance to the lens from
%\begin{equation}
%    D_{\rm L} = \left(\theta_{\rm E}\pi_{\rm E} + \frac{1}{D_{\rm s}} \right)^{-1}.
%\end{equation}
\begin{table*}[t]
    \centering
    \caption{Positions, proper motions and $RUWE$ \KR{(see footnote \ref{foot:RUWE})} parameters for all the events in the analyzed sample, taken from the GDR3 catalog. In the last two columns we also provide the alert name, for those detected through Gaia Science Alerts (GSA), and the GDR3 source identifier.}
    \begin{tabular}{cccccccc}
        \hline
        \hline
        \noalign{\smallskip}
        Event & RA (J2016)& Dec (J2016)& $\mu_{RA^*}$ [mas/yr]& $\mu_{Dec}$ [mas/yr]& $RUWE$ & GSA name & GDR3 \verb|source_id| \\
%        &(J2000)&(J2000)&[mas/yr]&[mas/yr]&\\
        \noalign{\smallskip}
        \hline
        \noalign{\smallskip}
        OB150145&270.17782&-35.15408&$-1.29 \pm 0.07$&$-6.51 \pm 0.05$&1.03&-&4041998223399082752\\
        OB150149&270.28819&-32.55773&$-2.01 \pm 0.21$&$-5.03 \pm 0.13$&2.35&-&4042928139682133120\\
        OB150211&262.35909&-30.98178&$-3.50 \pm 0.29$&$-7.22 \pm 0.21$&1.05&-&4058004814930630912\\
        OB160293&268.16140&-32.48960&$-1.49 \pm 0.09$&$-7.48 \pm 0.07$&0.92&-&4043504794840743040\\
        OB160689&261.14944&-30.13245&$-1.80 \pm 0.36$&$-7.25 \pm 0.23$&1.38&-&4059051309459806208\\
        OB180410&272.22770&-27.16974&$-4.51 \pm 0.16$&$-6.84 \pm 0.11$&1.14&Gaia18cho&4063011505688647296\\
        OB180483&262.64177&-27.49183&$4.85 \pm 1.23$&$-2.13 \pm 0.74$&1.34&Gaia18ayh&4061439448723558016\\
        OB180662&266.87754&-32.52442&$-5.74 \pm 0.59$&$-6.09 \pm 0.34$&1.13&Gaia18cej&4054012488194100096\\
        OB190169&265.98559&-32.87095&$-0.87 \pm 0.11$&$-4.54 \pm 0.06$&0.97&Gaia19drv&4054032245075925760\\
        \noalign{\smallskip}
        \hline
        \hline
    \end{tabular}
    \label{tab:Gaia_parameters}
\end{table*}
%The distance to the source $D_{\rm s}$ is also an additional parameter, that cannot be derived from the light curve model.

In principle, we could assume fixed distance to the source in our calculations - all the events analyzed here lie towards the galactic center, and so it is expected that the sources belong to the bulge population, particularly when their location on the CMD coincides with the Red Clump (see Figure \ref{fig:multipanel_cmd}).
Nonetheless, we decide to weight them with the stellar density distribution, which gives more realistic results. We also decide not to use parallax measurements from GDR3, nor the distance estimates based on them \citep{Bailer-Jones2021}.  The reason is that the sources mostly lie in the bulge, and so Gaia parallax is not measured accurately enough (with the signal of the order of $\sim 0.1~\rm mas$), especially in such crowded fields. Indeed, after inspecting the parallax values quoted in GDR3, we found that the measured parallax is either negative, or \verb|parallax_over_error| $\lesssim 1$, meaning that these measurements do not carry useful information.

It is worth noting that in principle mass and velocity distributions for stellar remnants are different from those of stars. Nonetheless, in this experiment we do not know \textit{a priori} if the lenses belong to the stellar remnant population, and so we cannot assume that from the beginning and use priors for black holes/neutron stars. Instead, we use a Galactic model and the mass function based on stars, and after assessing the amount of light to the lens, we examine the scenario of it being a ``regular" star. We note, that if the velocity of a lens is higher, as expected for NSs (e.g. \citealt{Hobbs:2005}) and some BHs (e.g. \citealt{Repetto:2017}), then the derived mass can be regarded as a lower limit\footnote{\KR{It is important to note that in microlensing we access only transverse velocity - the natal-kick velocities of such objects can in principle have any direction and so, due to projection, they will not necessarily be seen as high velocity lenses.}}.

\begin{figure}
    \centering
    \includegraphics[width=0.4\textwidth]{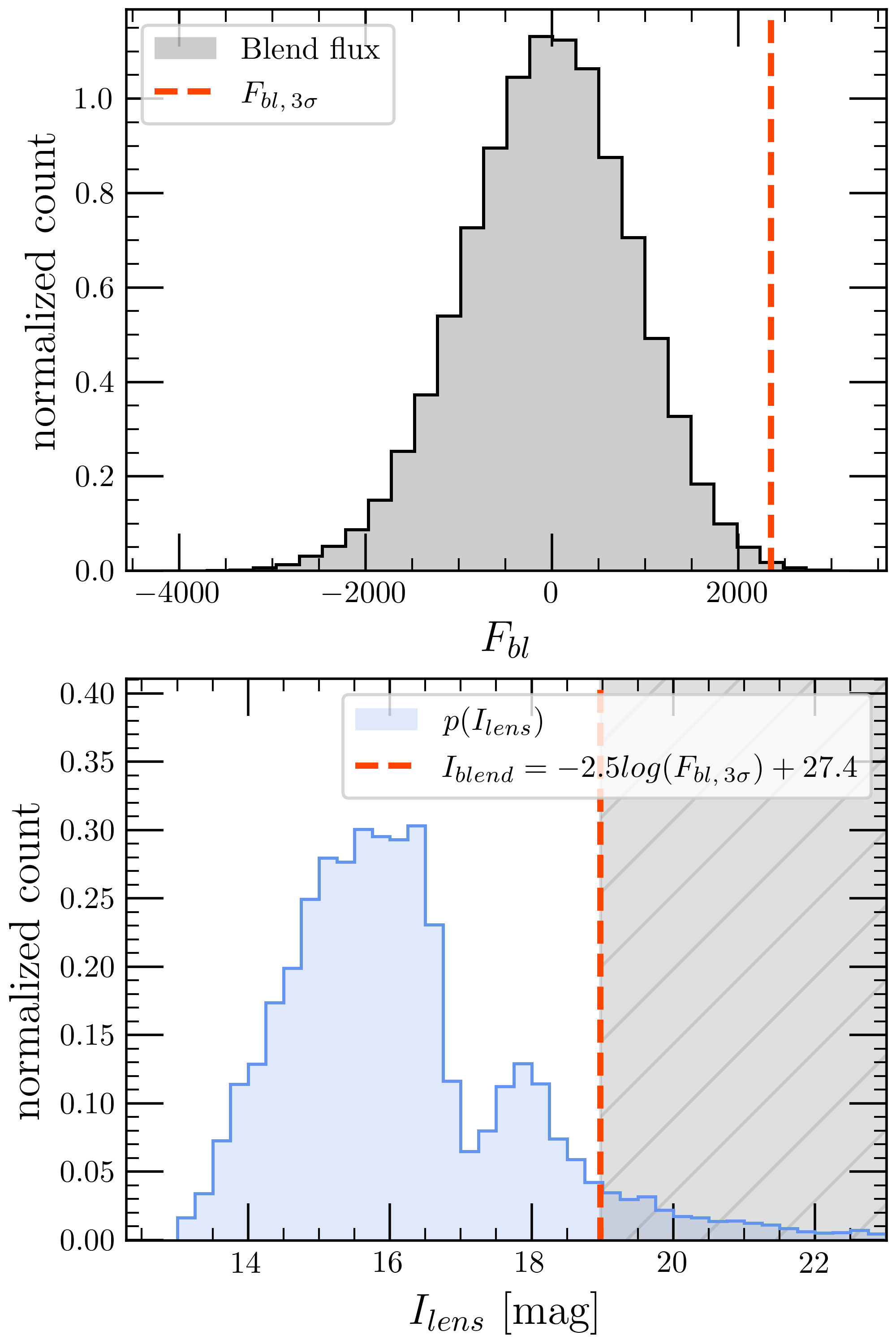}
    \caption{Visualization of the $p_{dark}$ calculation procedure for the example event \twoninethree. \textit{Top panel:} posterior distribution of the blend flux (in arbitrary units) \KR{resulting from the light curve modeling}. Red line shows the blend flux limit we assume for the $p_{dark}$ calculations. \textit{Bottom panel:} posterior distribution $p(I_{lens})$ of the lens brightness $I_{lens}$ (see details in the text). Red line shows the \KR{upper} limit on the blend brightness derived from the top histogram. All the samples for which the lens is brighter than the limit (to the left \KR{from the red line}) contribute to the dark lens scenario.}
    \label{fig:pdark_hist}
\end{figure}

\subsection*{Lens light}
To claim that a lens is a stellar remnant, not only the mass, but also the blend flux has to be investigated.
%Here we again follow the strategy presented in \cite{Wyrzykowski2016} and \citealt{Mroz+Wyrzykowski2021}.
After constructing the posterior distributions for the lens mass and distance, one can estimate the brightness $I_{lens}$ that the object of these properties should emit, under the assumption that it is a Main Sequence (MS) star. For this purpose we use the empirical mass-luminosity relation found by \cite{Pecaut+Mamajek2013}\footnote{\url{http://www.pas.rochester.edu/~emamajek}}. \KR{If we denote this relation by $L(M_{\rm L})$, we can define $I_{lens}$ as
\begin{equation}
    I_{lens} = -2.5\log{\frac{L(M_{\rm L})}{4\pi D_{\rm L}^2}} + A_{lens},
\end{equation}
where $A_{lens}$ is extinction to the lens.}

To estimate $A_{lens}$, we use extinction maps from \cite{Nataf.2013.A} for the 4 events lying in the OGLE-III fields covered in their analysis.
For the remaining five events we derive the extinction value based on the Red Clump position on the CMD, using similar procedure as Nataf et al., using their de-reddened RC brightness.
%For the case of \twooneone\, there is no OGLE $V$-band measurements, and so we calculate the RC position based on 
The $A_{lens}$ parameter is the extinction integrated along the whole distance to the source, so we treat it as an upper limit on the lens extinction. Again, for the sake of claiming if the lens is a stellar remnant or not, this is a conservative assumption. In addition, this simplification is justified by the fact that most of the disk dust between the observer and the source resides within the first few kiloparsecs from Earth, especially for the events with larger galactic latitudes.

\KR{Having a distribution of $I_{lens}$, one can compare it with the brightness of the blend,
%which is calculated based on the blending parameter $g$ and source flux $F_{\rm s}$ obtained from the light curve model. 
which is one of the products of the light curve modeling.
%Blend flux $F_{bl}$ is the sum of contributions from another star (or group of stars) that happens to lie within the PSF. It might also be due to companions, either to the source star, to the lens, or to both.
In the case where the total blend brightness is higher than $I_{lens}$, we get more light from the blend than is expected from the lens of given mass at a given distance. Such scenario is very common and easy to explain, as the excess can be attributed to any other sources lying on the same line of sight and not participating in microlensing - the lens is not the only light source contributing to the blending light, which is common for the galactic bulge direction. On the other hand, if the blend brightness calculated from the photometric model is lower than $I_{lens}$, the situation is opposite - there is not enough light emitted by the blend to explain a MS star of given mass at given distance. In other words, such MS star would be too bright compared to the expected blend light\footnote{\KR{We treat MS stars as a reference in this analysis, but more evolved star of given mass would be even brighter}}. It suggests that the lensing object is in fact not luminous. Then, if it is massive enough, it is considered a candidate for a stellar remnant.}

%The distribution of blend flux $F_{bl}$ that is obtained from the light curve modeling can be wide and can have part of it on the negative side of blend flux. While small levels of negative blending are acceptable, as it might be the result of the increased background signal due to the crowding in the bulge, it needs careful treatment when determining if the lens is dark or not. The reason is that in principle 
\KR{To perform the comparison between $I_{lens}$ and blend brightness, the latter has to be estimated based on the blend flux distribution $F_{bl}$, that is the product of the light curve modeling. To be conservative, we choose a 3-$\sigma$ upper limit on the blend flux (red dashed line on the top panel of Figure \ref{fig:pdark_hist}). To compare it with $I_{lens}$ we translate this limit to magnitudes and call it $I_{blend}$ (red dashed line on the bottom panel of Figure \ref{fig:pdark_hist}).}

\KR{As mentioned above, the lens is expected to be dark for each set of parameters (each link of MCMC chains) resulting in $I_{lens} < I_{blend}$. To formally assess the probability $p_{dark}$ that the lens is dark, we integrate all the samples for which $I_{lens} < I_{blend}$:
\begin{equation}
    p_{dark} = \frac{ \int_{0}^{I_{blend}} p(I_{lens}) \,dI_{lens} }{ \int_{-\infty}^{\infty} p(I_{lens}) \,dI_{lens} },
\end{equation}
where $p(I_{lens})$ is the posterior distribution of the $I_{lens}$ brightness. We illustrate this procedure on the Figure \ref{fig:pdark_hist}: all the samples for which lens brightness can be explained by the blend are grayed-out on the bottom panel. For the remaining part of the distribution, the lens with given mass at given distance would be too bright to explain it with a MS star, and thus this region corresponds to the dark lens scenario.}
%This approach allows to properly take into account small levels of negative blending.

\subsection{Physical parameters - results}

\begin{figure*}
    \centering
    \includegraphics[width=1.1\textwidth]{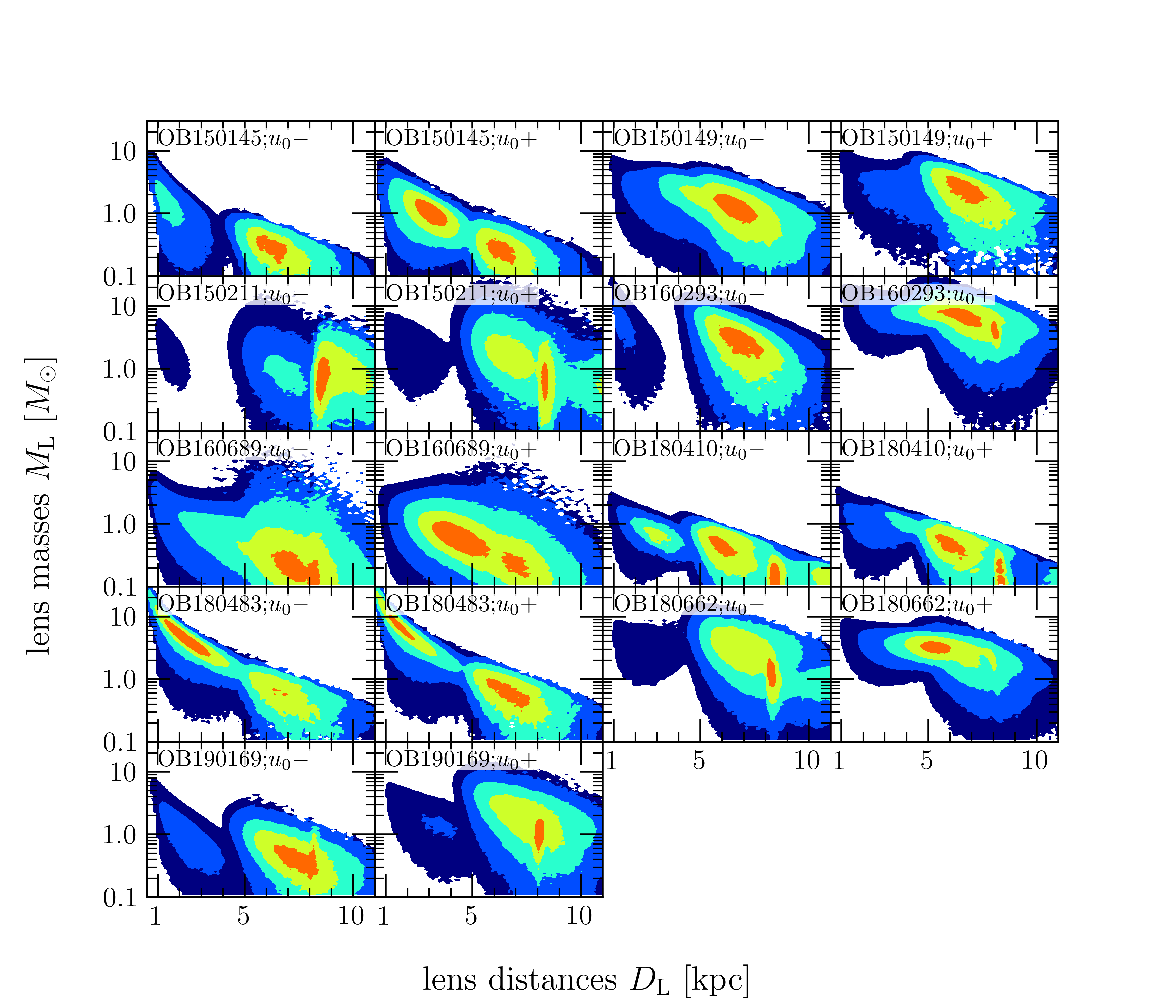}
    \caption{Posterior distributions for lens mass and distance, under the assumption of the Kroupa mass function (slope $\alpha=-2.35$) -- see text for the details. The orange, yellow, cyan, blue and dark blue contours mark 1,2,3,4 and 5--$\sigma$ confidence levels, respectively. Note, that we show the distributions for all the possible solutions (both $u_0+$ and $u_0-$ models for each event), even though some of them can be rejected based on the light curve analysis. We list the results for preferred solutions in Table \ref{tab:DLC_results}.}
    \label{fig:DLC_results}
\end{figure*}
\begin{table*}
  \centering
  \caption{Results of Bayesian analysis in the form of median values of marginalized distributions for lens masses, distances and Einstein radii, along with the probabilities that the lens is dark. The error values reflect the 68\% confidence level intervals. We also show the 2--$\sigma$ lower limit on the lens mass $M_{\rm L, 2\sigma min}$. For each event we show the more preferred solution (either $u_0+$ or $u_0-$) and results for the two priors on the mass function (see text for more information.)}
  \label{tab:DLC_results}

     \begin{tabular}{cccccccc}
        \hline
        \hline
        \noalign{\smallskip}
        Event&Model & Mass prior & Lens mass $M_{\rm L}$ [$M_{\odot}$] &$M_{\rm L, 2\sigma min}$ [$M_{\odot}$] & Lens distance $D_{\rm L} $ [kpc]  &  $\theta_{\rm E}$ [mas]& $p_{\rm dark}$ \\
        \noalign{\smallskip}
        \hline
        \noalign{\smallskip}
        OB150145 & $u_0-$ & Flat & $0.83^{+3.49}_{-0.53}$ & 0.17 & $2.14^{+4.17}_{-1.38}$ & $1.00^{+4.26}_{-0.63}$ & 49.9\% \\
        & $u_0-$ & Kroupa MF & $0.30^{+0.17}_{-0.12}$ & 0.09 & $6.15^{+0.83}_{-0.77}$ & $0.36^{+0.19}_{-0.14}$ & 5.8\% \\
        \noalign{\smallskip}
        OB150149 & $u_0-$ & Flat & $1.80^{+0.83}_{-0.64}$ & 0.70 & $6.00^{+0.93}_{-1.14}$ & $0.97^{+0.42}_{-0.32}$ & 92.5\% \\
        & $u_0-$ & Kroupa MF & $1.40^{+0.75}_{-0.56}$ & 0.45 & $6.35^{+0.95}_{-0.95}$ & $0.77^{+0.38}_{-0.29}$ & 79.0\% \\
        \noalign{\smallskip}
        OB150211 & $u_0+$ & Flat & $2.74^{+3.35}_{-1.65}$ & 0.56 & $6.80^{+1.47}_{-1.07}$ & $1.05^{+0.72}_{-0.61}$ & 28.4\% \\
        & $u_0+$ & Kroupa MF & $1.04^{+1.41}_{-0.49}$ & 0.30 & $8.20^{+0.17}_{-1.75}$ & $0.43^{+0.57}_{-0.16}$ & 4.7\% \\
        \noalign{\smallskip}
        OB160293 & $u_0-$ & Flat & $4.02^{+1.86}_{-1.48}$ & 1.46 & $6.23^{+0.86}_{-0.70}$ & $1.21^{+0.54}_{-0.44}$ & 98.9\% \\
        & $u_0-$ & Kroupa MF & $2.98^{+1.75}_{-1.28}$ & 0.85 & $6.58^{+0.91}_{-0.80}$ & $0.91^{+0.52}_{-0.39}$ & 93.6\% \\
        \noalign{\smallskip}
        OB160689 & $u_0+$ & Flat & $1.10^{+0.79}_{-0.48}$ & 0.31 & $4.00^{+1.60}_{-1.00}$ & $1.11^{+0.50}_{-0.42}$ & 42.2\% \\
        & $u_0+$ & Kroupa MF & $0.68^{+0.54}_{-0.36}$ & 0.13 & $4.48^{+2.10}_{-1.19}$ & $0.79^{+0.48}_{-0.40}$ & 16.2\% \\
        \noalign{\smallskip}
        OB180410 & $u_0-$ & Flat & $0.58^{+0.24}_{-0.21}$ & 0.16 & $5.62^{+0.89}_{-2.53}$ & $0.59^{+0.45}_{-0.22}$ & - \\
        & $u_0-$ & Kroupa MF & $0.42^{+0.24}_{-0.26}$ & 0.08 & $6.09^{+2.23}_{-0.91}$ & $0.42^{+0.30}_{-0.25}$ & - \\
        \noalign{\smallskip}
        OB180483 & $u_0-$ & Flat & $6.82^{+5.56}_{-2.63}$ & 2.55 & $1.55^{+0.71}_{-0.62}$ & $5.35^{+4.34}_{-2.04}$ & 99.6\% \\
        & $u_0-$ & Kroupa MF & $4.65^{+3.12}_{-2.08}$ & 0.47 & $2.07^{+1.03}_{-0.69}$ & $3.68^{+2.42}_{-1.62}$ & 92.3\% \\
        \noalign{\smallskip}
        OB180662 & $u_0+$ & Flat & $3.33^{+0.65}_{-0.60}$ & 2.02 & $5.28^{+0.80}_{-0.66}$ & $1.48^{+0.25}_{-0.23}$ & 99.9\% \\
        & $u_0+$ & Kroupa MF & $3.15^{+0.66}_{-0.64}$ & 1.54 & $5.35^{+0.95}_{-0.67}$ & $1.42^{+0.25}_{-0.26}$ & 99.3\% \\
        \noalign{\smallskip}
        OB190169 & $u_0-$ & Flat & $0.71^{+0.41}_{-0.28}$ & 0.25 & $6.63^{+1.26}_{-0.91}$ & $0.54^{+0.25}_{-0.20}$ & 0.7\% \\
        & $u_0-$ & Kroupa MF & $0.50^{+0.32}_{-0.22}$ & 0.15 & $7.12^{+1.01}_{-1.03}$ & $0.40^{+0.22}_{-0.17}$ & 0.1\% \\
        \noalign{\smallskip}                
        \noalign{\smallskip}
        \hline
        \hline
     \end{tabular}
\end{table*}

Below we provide results of the Bayesian analysis for each event separately.

\subsubsection*{\onefourfive}

In the Figure \ref{fig:DLC_results} we present the Bayesian analysis results for both the $u_0-$ and $u_0+$ solutions, although the $u_0+$ one is virtually excluded by the light curve analysis (see Section \ref{sec:OB150145_lc}). The $M_{\rm L}$ vs $D_{\rm L}$ distribution shows a bi-modality as lens is either in the disk or in the bulge. This overall structure is visible in most of the events in the sub-sample, although usually a bulge lens is the preferred solution, as expected. Indeed, for the $u_0-$ solution here, the bulge scenario is much more preferred and it yields a low lens mass $M_{\rm L}=0.30^{+0.17}_{-0.12}$ at $D_{\rm L}=6.12^{+0.87}_{-0.75}$, with very low chance of being dark (see Table \ref{tab:DLC_results}).

\subsubsection*{\onefournine}

In the $M_{\rm L}-D_{\rm L}$ plane, the two solutions behave in a similar way, with only one maximum each. The preferred $u_0-$ solution places the lens at around 6.3 kpc with mass in the range $0.7-2~M_{\odot}$. The 79\% probability of being dark suggests that the lens might be a massive white dwarf or a neutron star.
This is the only event in the sample that has $RUWE$ parameter value substantially higher than one (see Table \ref{tab:Gaia_parameters}) and so we decided to perform the Bayesian analysis with the source proper motion prior (taken from GDR3 and thus, given high $RUWE$, unreliable) loosened. Nominally we use 1--$\sigma$ value from GDR3 as a width of the prior. Here we are broadening the prior to ten standard deviations, as the Gaia values should not be fully trusted. We do not observe significant change in the $M_{\rm L}-D_{\rm L}$ plane, apart from the anticipated broadening of the posterior distributions.

\subsubsection*{\twooneone}
The $M_{\rm L}-D_{\rm L}$ plane is again similar for the two solutions, with only the bulge lens scenario being viable. In both distributions one can see an additional, sharper structure at around 8 kpc. Loosening the Gaia prior makes it merge with the wider bulge distribution, which means it is caused by the source proper motion prior. Similar structure is visible in some of the other events in the sub-sample.

\subsubsection*{\twoninethree}
\label{sec:OB160293_MLDL}

The lens in this event is one of the best candidates for a massive remnant in the analyzed sub-sample. Multiple factors in the light curve modeling suggest that the $u_0-$ model is correct (see \ref{sec:OB160293_lc}) and so we present values yielded by this model in Table \ref{tab:DLC_results}.

Although $\theta_{\rm E}$ does not have extreme value, thanks to a small microlensing parallax, the resulting lens mass is somewhat large and is expected to lie in the range $1.7-4.9~M_{\odot}$ for bulge lenses. This results in a 94\% probability for the lens being dark. Hence, it is an excellent candidate for a neutron star or a stellar mass black hole. There is a possibility that the lens lies in the disk, which yields even higher masses, but it is strongly disfavored in our analysis (see relevant panel in Figure \ref{fig:DLC_results}).

\subsubsection*{\sixeightnine}
\label{sec:OB160689_MLDL}

The blending for this event is not well constrained from the photometric model, but is likely non-negligible (see Section \ref{sec:OB160689_lc}). As a result, the proper motion detected by Gaia is a combination of source and lens proper motion, which means that we can not use it directly in the prior as source proper motion.
%Instead, we draw the source proper motion from a wide gaussian distribution, centered at the value provided in GDR3 but with the width extended by a factor of ten.
\KR{Instead, when calculating the prior on relative proper motion, we assume the source proper motion to be $(\mu_{l}, \mu_{b}) = (-6.12, -0.19)\pm 2.64\,\rm mas/yr$ \cite{Schonrich2010}}. This corresponds to a typical motion of the galactic center relative to the Sun and is a reasonable assumption as the source most likely resides in the bulge. The same procedure was applied in \cite{Mroz2021}.
Naturally, the resulting $M_{\rm L}$ vs $D_{\rm L}$ distribution is much wider compared to the one resulting from the approach with the Gaia value. Because of the low expected lens mass and presumably strong blending, it is very unlikely that this event hosts a remnant lens.
%Nonetheless, it is skewed toward higher masses and, even though the blending is significant, the probability of lens being dark is very high. It suggests that the blending light does not come from the lens, but rather from a star which position coincides with the Earth-source line of sight.
%High resolution follow-up (e.g. with Adaptive Optics facilities) would be helpful to , which would allow to confirm the dark lens scenario and further constrain its parameters.

\subsubsection*{\fouronezero}

The $u_0-$ solution is preferred in terms of goodness of fit in the light curve analysis ($\Delta \chi^2\approx19$). Redoing the photometric fit with the blending fixed to zero helps to resolve part of the tension between the microlensing parallax solutions (see Section \ref{sec:OB180410_lc}), and so we use this model to estimate the physical parameters. With the blending parameter fixed to zero we can not assess the probability of the lens to be dark, as it is assumed to be dark in the first place. Nonetheless, the mass of the lens yielded by our analysis is very low, and so it is most likely an ordinary dwarf star lying in the bulge.

\subsubsection*{\foureightthree}
There is a clear bi-modality in the $M_{\rm L}$ vs $D_{\rm L}$ distribution due to the duality in the possible lens populations. Each of the $u_0+$ and $u_0-$ solutions have the disk lens solution, in which the lens is relatively heavy, in the range $5-10~M_{\odot}$, and bulge lens solution where $M_{\rm L} \lesssim 1~M_{\odot}$. The $u_0+$ case is excluded with the flux Spitzer limits, so we know $u_0-$ the correct one. From the posterior distribution of  the lens mass and distance for the negative solution, it seems that the scenario of heavier lens located in the disk is more preferred, although there is a small region of similar probability for the bulge lens case. As the blending is very low for this event, the lens is a good remnant candidate with estimated mass $M_{\rm L} = 4.65^{+3.12}_{-2.08}$ and $p_{dark}\approx92\%$.

\subsubsection*{\sixsixtwo}

The positive solution is favored both in terms of goodness of fit and compatibility of ground-based-only and Spitzer-``only'' parallax solutions (see section \ref{sec:OB180662_lc}). The expected mass for this event remains somewhat large for a range of lens distances (see Figure \ref{fig:DLC_results}). Additionally, the blending level is very low, which makes this event one of the best candidates in our sub-sample to host a dark remnant with $M_{\rm L} = 3.15\pm0.65$ and $p_{dark}\approx 99\%$.

\subsubsection*{\onesixnine}

In the $M_{\rm L}-D_{\rm L}$ plane, in both of the solutions we can see the impact of the GDR3 source proper motion prior, similarly to \twooneone\, - the narrow structure gradually disappears with the increase of the width of the prior. None of the solutions shows the prospect for the remnant lens, and the preferred $u_0-$ case suggests particularly low probability for that with mass of the lens $M_{\rm L}\approx 0.5~M_{\odot}$ and $p_{dark}<1\%$.
%In the light curve of \onesixnine,\, at $HJD'\approx 8720$, there is a clear deviation from the light curve model lasting $\sim 5$ days. It is not an instrumental effect as it appears in both OGLE and KMTNet data. While the scenario of the deviation being a planetary anomaly should not be completely excluded, there is a clear, irregular source variability throughout the light curve, with the amplitude of some peaks comparable to the "anomaly". Thus, we conduct the analysis using the single lens model and attribute the feature to the variability of the source. We note that one of the Gaia epochs was taken during the "anomaly". In an unlikely scenario of the deviation being due to a planet, it might be an interesting point of the analysis of the Gaia astrometric data.

\section{Gaia predictions}
\label{sec:gaia_predictions}
\begin{figure*}
    \centering
    \includegraphics[width=\textwidth]{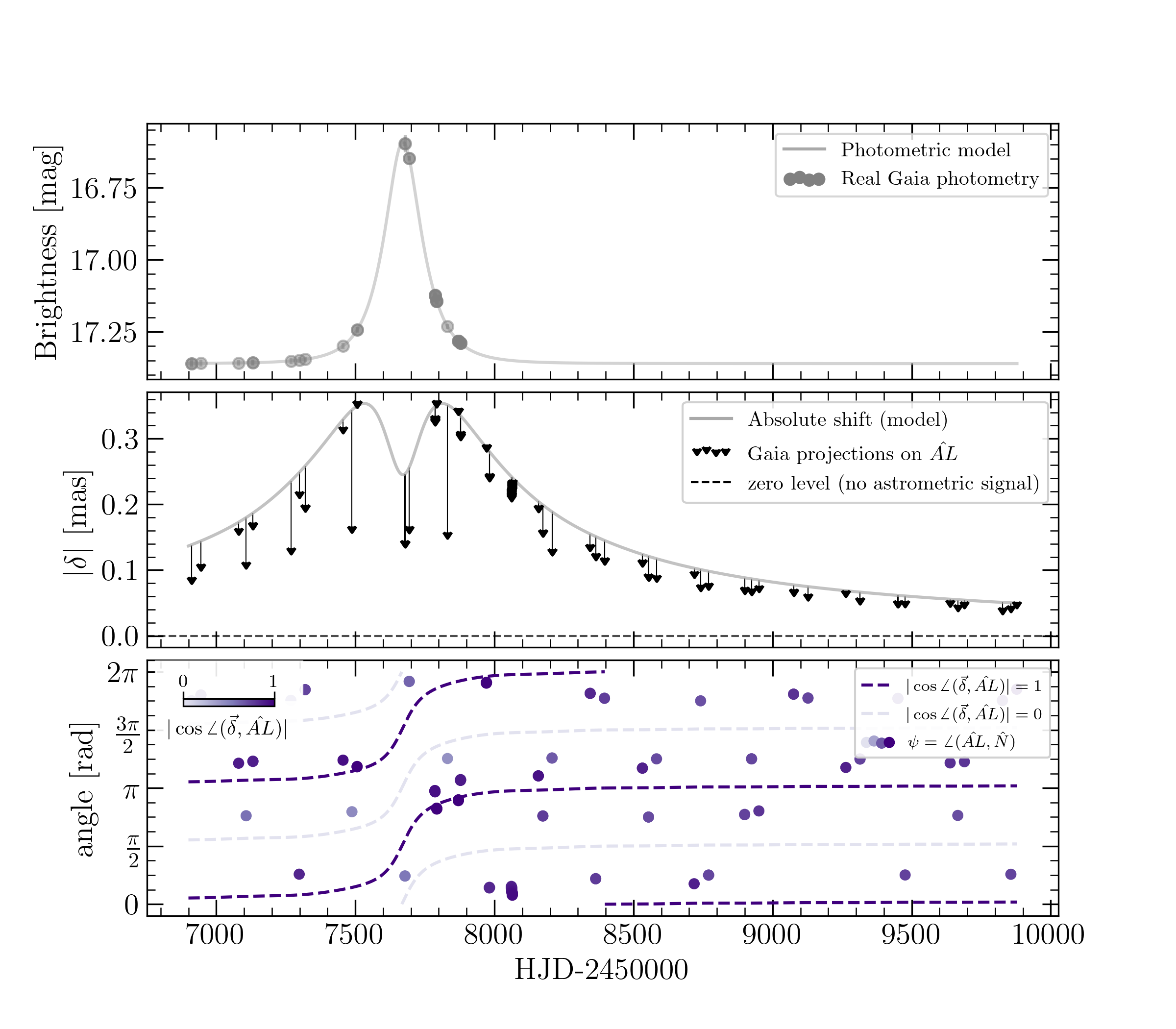}
    \caption{Visualization of all the known information about the Gaia measurements, under the assumption of the photometric microlensing model \KR{and $\theta_{\rm E}=0.91\,\rm mas$ (see Table \ref{tab:DLC_results})}, for the case of \twoninethree. \textit{Top panel:} Photometric model with marked Gaia measurements. \textit{Middle panel:} Absolute astrometric shift calculated based on the photometric model (gray line). The black arrows indicate how much the astrometric signal is ``reduced" due to the projection on the $\widehat{AL}$ direction. \textit{Bottom panel:} Points represent scanning angles of Gaia mission for this field. Dark dashed lines show the direction of the astrometric shift vector $\vec{\delta}$ $\pm180^{\circ}$, while light-colored dashed lines mark perpendicular directions.
    \KR{Gaia only provides measurements along the instantaneous scanning direction $\widehat{AL}$. As a result, whenever astrometric shift vector $\vec{\delta}$ and $\widehat{AL}$ are aligned or counter-aligned ($|\cos{\angle{(\vec{\delta},\widehat{AL})}}| \approx 1$), the signal measured by Gaia is larger. This occurs when the points on the bottom panel are darker and situated closer to one of the dark lines. Indeed, one can see that darker points on the bottom panel correspond to shorter arrows on the middle panel, meaning that more astrometric signal is ``available'' for Gaia.}
    %The closer the points are to the dark lines, the more aligned (or counter-aligned) the shift direction and scanning direction are, which results in larger signal measured by Gaia, as it is only sensitive in $\widehat{AL}$ direction. One can see that whenever $|\cos{\angle{(\vec{\delta},\widehat{AL})}}| \approx 1$ (and so the dots are closer to one of the dark dashed lines), the arrows in the middle panel are shorter, which means that most of the signal is measured by Gaia. Conversely, when scanning direction and microlensing shift vector are close to perpendicular configuration, the measured displacement is ``reduced", and thus respective arrows are longer.
    }
    \label{fig:OB160293_astroulensing_model}
\end{figure*}
All the events analyzed here were chosen under the condition that they had been observed by the Gaia mission and so there will be high precision astrometry available for them in the Gaia Data Release 4 (GDR4, to be published $\sim$2026). Knowing the epochs of astrometric measurements, the expected uncertainties and assuming $\theta_{\rm E}$ values derived in Section \ref{sec:physical}, we perform realistic simulations of the Gaia astrometry and investigate the prospects for measuring $\theta_{\rm E}$ and consequently the masses of the lenses in the studied events.

The Gaia mission will provide 2D astrometry only for the brightest objects ($G>13$ mag). For the remaining ones, only one-dimensional astrometry, measured along the instantaneous scanning direction $\widehat{AL}$, which we will denote with angle $\psi$ measured from the North direction eastward. In the simulations, we use actual values of scanning angles and epochs calculated based on the Gaia scanning law, taken from GOST\footnote{\url{https://gaia.esac.esa.int/gost/}} (Gaia Observation and Forecast Tool). We treat each Gaia visit as a single epoch, without dividing into sub-measurements from each AF (Astrometric Field)\footnote{Gaia constantly rotates, and so whenever an object is observed, it transits the focal plane, passing through (nominally) 9 AFs.}. To estimate the error bars, we follow the conservative approach of \cite{Rybicki2018}, where the centroiding errors from \cite{deBruijne2014} are increased by 50\% to account for potential systematics. They also take into consideration the fact that there are measurements from multiple AFs within one epoch, which scales down the error bars by a factor of $\sqrt{9}=3$.

\subsection*{Centroid trajectory model}

The shift of the centroid position from the position of the source can be expressed as (e.g. \citealt{Dominik2000})
\begin{equation}
    \vec{\delta} = \frac{\vec{u}}{u^2+2}\theta_{\rm E}.
    \label{eq:centroid_shift}
\end{equation}
Because we are investigating the Gaia potential to detect the astrometric microlensing effect, for now we disregard the source proper motion and parallax. In simulated data they will be  easily distinguished from the microlensing and thus, to first order, should not affect how well the microlensing signal can be recovered. On the other hand, while dealing with the real data, potentially contaminated by unknown systematics, a full model with parallax and proper motions will have to be applied.
\begin{figure*}
    \centering
    \includegraphics[width=\textwidth]{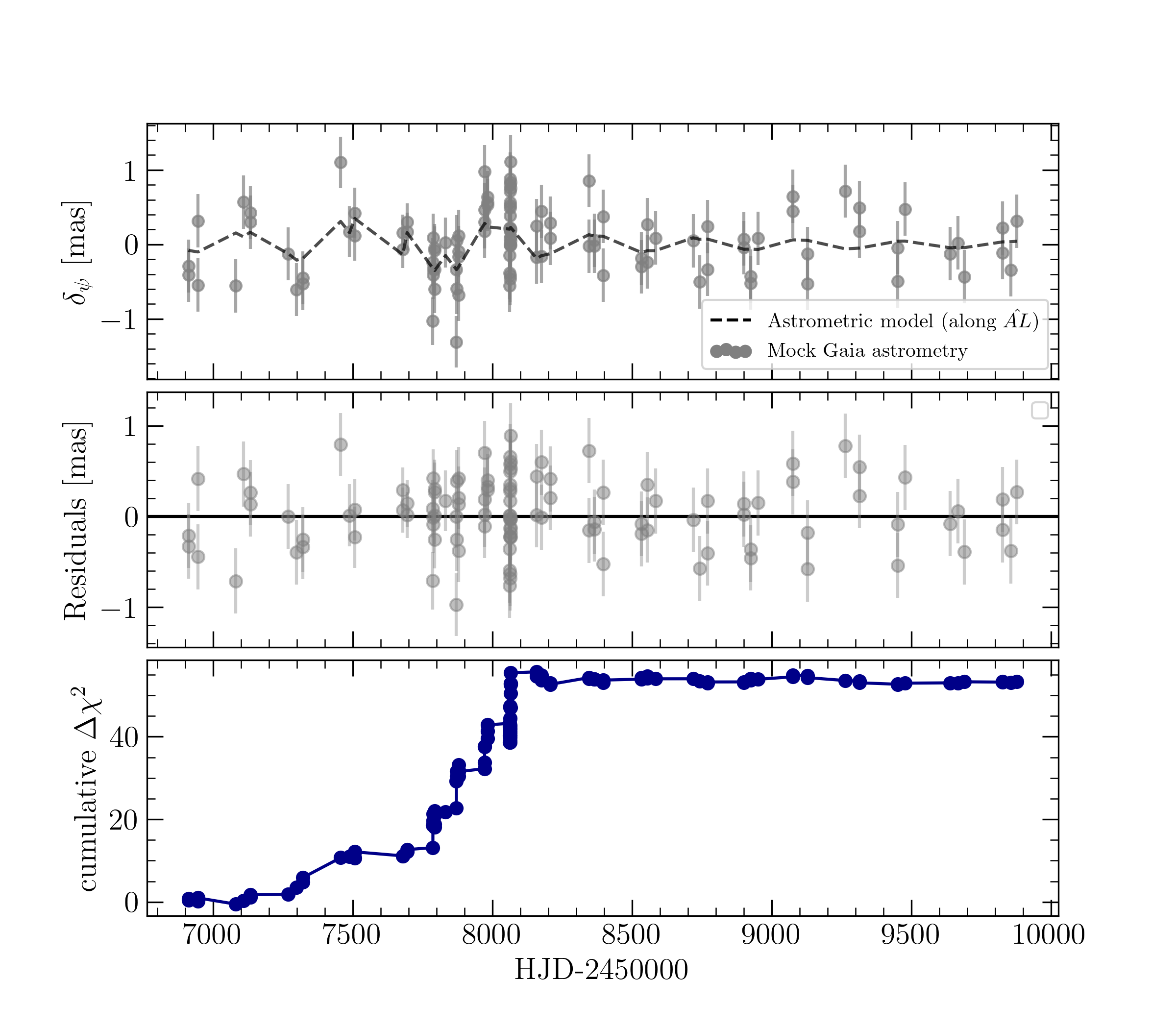}
    \caption{The cumulative plot of $\Delta\chi^2$ of the astrometric 1-D fit for the event \twoninethree\, for an example realization of the Gaia data \KR{and $\theta_{\rm E}=0.91\, \rm mas$ (see Table \ref{tab:DLC_results})}. \textit{Top panel:} Mock Gaia astrometry (gray points) of the centroid displacement during microlensing (proper motion and parallax not included here). Model used to generate the data is marked by the dashed line. \textit{Middle panel:} Residuals of the data and the model shown on the top panel. \textit{Bottom panel:} Cumulative $\Delta\chi^2$, which is the difference between the null model and the astrometric microlensing model (see details in the text).}
    \label{fig:OB160293_astroulensing_data}
\end{figure*}

Based on the Equation \ref{eq:centroid_shift}, one can see that information about the astrometric shift comes from three somewhat distinct instances: the relative separation $u(t)$, the direction of the lens-source relative motion, and the Einstein radius. The first is almost always available as the relative separation is a function of parameters easily derivable from the standard light curve, namely $(t_0, u_0, t_{\rm E})$. It means that, for all standard events that are relatively well covered, one can already predict the astrometric shift of the centroid for each solution (which traces out an ellipse, see e.g. \citealt{Dominik2000} for details), although without orienting it on the sky or scaling it to the absolute angular units. For events considered in this analysis, we also have very strong constraints on the microlensing parallax vector, which in turn gives the direction of the lens-source relative proper motion (see Equation \ref{eq:piE_direction}). In consequence, we are able to align the centroid shift trajectory on the sky only using the light curve model. As a result, for each photometric solution, the only information that needs to be recovered from the astrometric data is $\theta_{\rm E}$ -- the scaling factor of the microlensing ellipse, whose shape and orientation are already known. It leads to the conclusion, that even astrometric data of seemingly insufficient precision, might provide enough information to determine the Einstein radius.

To construct the model of the astrometric microlensing shift as seen by \Gaia, one has to project the vector $\vec{\delta} = (\delta_{N}, \delta_{E})$ 
%Knowing the North and East components of the centroid shift $\vec{\delta} = (\delta_{N}, \delta_{E})$ from the light curve, one has to project it
onto the scan direction $\widehat{AL}$, and so the one-dimensional astrometric microlensing signal observed by the satellite can be written as:
\begin{equation}
    \delta_{\psi} = \vec{\delta}\cdot\widehat{AL} = \delta_{N}\cos{\psi} + \delta_{E}\sin{\psi}.
\end{equation}

In the Figure \ref{fig:OB160293_astroulensing_model}, we visualize the ``transition" from the two-dimensional signal to the one-dimensional Gaia data. The absolute astrometric microlensing shift presented in the middle panel (gray continuous line) is ``reduced" by the projection on the $\widehat{AL}$ direction, which is denoted by arrows. The level of this reduction depends on the angle between the astrometric shift and the scanning direction, which changes in time, as shown on the bottom panel of the figure.

We use posterior distributions on the physical parameters derived in Section \ref{sec:physical} to calculate $\theta_{\rm E}$ and generate 1-D Gaia astrometry for all the events in the sub-sample.
\subsection{Detectability}
\label{sec:gaia1D-detectability}

To quantify the detectability of astrometric microlensing, we first simulate the Gaia data, following steps and assumptions from the previous subsection. Then we calculate the $\chi^2$ statistic for the ``null-model", which is the case where astrometric microlensing is not present (effectively $\theta_{\rm E} = 0$):
\begin{equation}
    \chi_{\rm null}^2 =  \sum_i \frac{(\delta_{\psi, i} - \delta_{\psi, 0, i})^2}{\sigma_i^2},
\end{equation}
where the sum is evaluated over all Gaia measurements and $\sigma_i$ are their respective uncertainties. Because we only consider the astrometric signal from microlensing, our null-model is simply the baseline level and thus $\delta_{\psi, 0} = 0$.
Finally, calculating the difference $\Delta\chi^2 = \chi_{\rm null}^2 - \chi^2$, between the $\chi^2$ of the null-model and the correct model with astrometric microlensing, allows us to evaluate the confidence level for detecting the astrometric signal present in the Gaia data.

%For $\Delta\chi^2$ calculations one can simulate the ``ideal" set of Gaia astrometry without scatter, which follows the model exactly or simulate the statistical scatter based on the expected uncertainties. The first approach allows to reliably claim what is the magnitude of the astrometric signal and is independent from particular realizations of the randomized, synthetic data sets.
%Neither of these approaches involve fitting of the model and thus potential problems (degeneracies, false positives, etc.) might be overlooked. To make sure that not only the strength of the astrometric signal is sufficient, but also the model can be reliably fitted to the data, we generate multiple sets of Gaia data and fit the astrometric microlensing model for each of the studied events.
In the top panel of the Figure \ref{fig:OB160293_astroulensing_data}, we show an example of one-dimensional astrometric Gaia data for \twoninethree, simulated with realistic scatter. As mentioned before, the only information that we need to extract from the astrometry is $\theta_{\rm E}$ - a scaling factor of the model whose shape is already known for a given set of photometric parameters. Thus, even though the visual inspection of the top panel of the Figure \ref{fig:OB160293_astroulensing_data} does not reveal any obvious signal, the bottom panel with the cumulative $\Delta \chi^2$ plot shows that (for event \twoninethree) the astrometric signal should be strong enough to detect it.
%The constraints coming from the light curve models are of course much stronger due to the difference in coverage and signal to noise. As a consequence, Gaia astrometry only realistically constraints value of $\theta_{\rm E}$ - the one parameter that light curve is not sensitive to.

%% figures:
%% 5-panel Gaia figure
%% try 6-panel Gaia figure (add chi2 plot)
%% modifications to the angle plot: segmented colormap, larger points, contours, add colors to absolute shifts plot, light curve Gaia pooints not grey
%% work on readibility of gaia 1D data
\subsection{Results}
\begin{figure}
    \centering
    \includegraphics[width=0.5\textwidth]{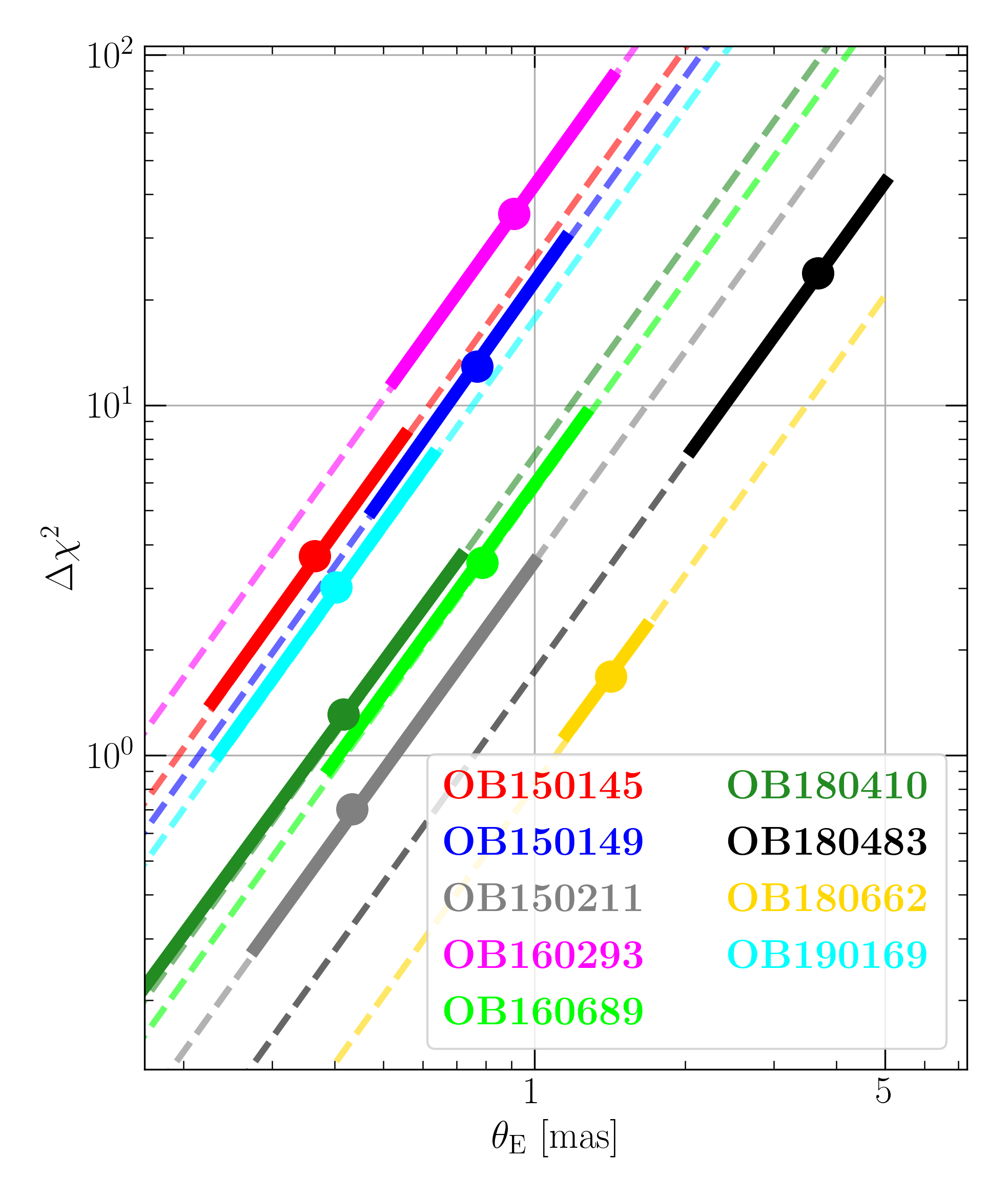}
    \caption{Detectability of the astrometric microlensing in Gaia data for the selected sub-sample of long timescale Spitzer events, plotted as a function of Einstein radius $\theta_{\rm E}$. The $\Delta\chi^2$ parameter is the difference between the null model and the best-fit astrometric microlensing model (see details in the text). Points represent expected $\theta_{\rm E}$ values for each event \KR{that were calculated in Section \ref{sec:mass_and_distance} and listed in Table \ref{tab:DLC_results}. Thickened lines mark 1--$\sigma$ errors on $\theta_{\rm E}$. We also simulated Gaia astrometric data for wider range of Einstein radii - the detectability for $\theta_{\rm E}$ in range from 0.1 to 5 mas is plotted with dashed lines.}}
    \label{fig:detectability}
\end{figure}
For all the events in the sub-sample, we assess the detectability of the astrometric microlensing signatures by comparing the null model to the model generated with the microlensing signal, as explained in the previous section. Two events require some caution in interpreting the results though. The first one is \sixeightnine, which is likely blended. Then, the signal measured by Gaia is a combination of the flux from the source and lens/blend(s). As a consequence, the astrometric microlensing signal expected here will be much weaker and the measurement much less robust. The second event is \onefournine. In that case the $RUWE$ parameter is notably larger than unity, which suggests that the Gaia 5-parameter astrometric solution can be unreliable. As a result, the real Gaia data for this event might look different than what we simulate here and so the measurement of astrometric microlensing signal may be more difficult.

We simulate the Gaia data and calculate $\Delta \chi^2$ for the nine events, using $\theta_{\rm E}$ values range spanning from 0.1 to 5 miliarcseconds, to construct a relation between $\theta_{\rm E}$ and $\Delta \chi^2$. Having statistical estimates of the lens masses and Einstein radii, presented in Section \ref{sec:mass_and_distance}, we are able to use this relation to determine for which events we expect to detect the astrometric microlensing signal in the Gaia data. The results are presented in Figure \ref{fig:detectability}, where we show the relations between $\Delta \chi^2$ and Einstein radius, also marking the $\theta_{\rm E}$ values that are expected from the Bayesian analysis \KR{(see Table \ref{tab:DLC_results})}. Out of the nine analyzed events, two of them show promising results in terms of Gaia astrometry detection capabilities: \twoninethree\, (magenta marker on Figure \ref{fig:detectability}) and \foureightthree\, (black). The difference between the astrometric null model and the microlensing model for the two events is $\Delta\chi^2_{\rm OB160293}\approx35$ and  $\Delta\chi^2_{\rm OB180483}\approx24$. Taking into account the one-parameter difference between the null-model and microlensing model (one $dof$ difference), it translates to about 6--$\sigma$ and 5--$\sigma$ expected detection level, respectively.

Generally, there are multiple factors impacting the detectability of the astrometric microlensing signal in the Gaia data. The most important ones are the Einstein radius, the brightness of the event, Gaia sampling and the scanning angles configuration. It is not a surprise that the two events mentioned above are preferred. In the case of \twoninethree\, all of the above requirements are met - in addition to the relatively large predicted $\theta_{\rm E}=0.91$ mas and several points covering the amplified part of the light curve, the event is bright having $I_{0,OGLE}=16.2$ mag. The latter is particularly important, because Gaia astrometric accuracy declines steeply with decrease in brightness. On the other hand, \foureightthree\, is faint with baseline at $I_{0,OGLE}\approx$18.6 mag (and even fainter in \Gaia\, as $I_{0,Gaia}\approx$20.0 mag), but we expect it to be detected in the Gaia data thanks to its extreme time scale $t_{\rm E}\approx330$ days, which translates into a large Einstein radius. Another favorable consequence of the long timescale is the fact that there are more data points from Gaia throughout the (significantly) amplified part of the light curve, which further enhanced detectability.

The second group consists of four events that are less likely to be detected, with the expected astrometric signal on the level of 1.5--3$\sigma$: \onefourfive\, (red marker on Figure \ref{fig:detectability}), \onefournine\, (dark blue), \onesixnine\, (cyan), and \sixeightnine\, (light green). The events from this group are bright and relatively long (see Table \ref{tab:fit_parameters}), but according to our mass/distance analysis, the most probable Einstein radius values for favored solutions are likely too small to be robustly detected by Gaia, even for such bright targets. The \onefournine\, event stands out here, but as already mentioned, it is more challenging to make predictions for due to the higher $RUWE$ value.

In the case of \twooneone\, (gray) and \fouronezero\, (green) the expected detectability is even smaller, with the astrometric microlensing signal at or below the 1--$\sigma$ level. The reason is that the Einstein radii are rather small ($\theta_{\rm E} \lesssim 0.5$ mas) and these events are fainter than the three targets mentioned before.

Finally, the mass measurement of the lens in the \sixsixtwo\, (yellow) event is also rather unlikely through the detection of astrometric microlensing in the Gaia data, even though the expected lens mass and Einstein radius are large (see Table \ref{tab:DLC_results}). In this case, the main factors affecting the detectability prospects are the relatively low brightness of the event ($I_{0,OGLE}=17.6$; additionally the source is very red, meaning the brightness in $G$-band is significantly lower) and duration of the event - even though the relatively short timescale $t_{\rm E}=66.6$ days translates to a somewhat large Einstein radius $\theta_{\rm E}=1.47$ mas, it results in fewer Gaia data points covering the event. Moreover, the scanning angle configuration is such that the 1-D astrometric signal is close to zero for a large fraction of the few important points. As a consequence, one of the best black hole candidates in our sample, with an expected dark lens of mass 2.5--4.2 $M_{\odot}$ is unlikely to be confirmed by the Gaia astrometry.

%\section{Discussion}
%\label{sec:discussion}
%\KR{Information about particular events from this section will be redistributed in other sections, mostly \ref{sec:massive_candidates}}
%Below we discuss the caveats related to the analysis of each of the event separately. We focus especially on the microlensing parallax measurements, as it has a critical impact on the mass of the lens estimates. We also discuss the results of the Bayesian analysis and posterior distributions of lens' masses and distances presented in Figure \ref{fig:DLC_results}.

\section{Summary and conclusions}
\label{sec:summary}

In this work, we analyzed a sub-sample of nine events chosen from the whole population of microlensing events observed by the \Spitzer\, Space Telescope in the years 2014-2019. The sub-sample was chosen from the events that had been also observed by the \Gaia\, mission, with long time scales and small microlensing parallaxes. The last two requirements were used to identify candidates for massive lenses.

Based on this small sub-sample, we demonstrate the procedures that will be applied to the whole Spitzer sample of microlensing events, which will allow us to populate $\lowercase{t}_{\rm E}$ -- $\pi_{\rm E}$ diagram and conduct statistical analysis of micrloensing parallax.
%The initial version of the plot, with the nine events analysed in this paper, is presented in Figure \ref{fig:piE-tE}.
Such analysis is necessary to complete the studies on the planets frequency in the Galaxy, which is based on microlensing planets detected in the \Spitzer\, campaign.

The detailed analysis of the selected sub-sample of nine events allowed us to identify candidates for black holes and neutron stars that can be later confirmed by the \Gaia\, time-series astrometry, which is expected to be released in the early 2026. Based on the Bayesian analysis incorporating the Galactic model and proper motion information from GDR3 (Section \ref{sec:mass_and_distance}) we found four candidates for dark remnant lenses: \twoninethree, \foureightthree, \sixsixtwo\, and \onefournine\, (see Table \ref{tab:DLC_results} for estimated masses). The masses of the lenses of the four candidate events lie somewhat on the edge of the known distinction between the neutron stars and stellar black holes. In the case of \onefournine\, it is most likely a neutron star. As for the remaining three candidates, the median mass suggests black holes, although the error bars are large and so the heavy neutron star scenario can not be excluded. The cases of \twoninethree\, and \foureightthree\,  are expected to have an astrometric microlensing signal detectable by Gaia, which in turn will allow to confirm the mass of the lens and their remnant nature.

%Recent advancements have expanded the scope of detection methods. Space telescopes, such as the James Webb Space Telescope, hold substantial promise for observing the effects of these remnants on neighboring celestial bodies. Additionally, X-ray and gamma-ray observatories can provide essential insights into their emissions and behavior.

%The quest to detect and weigh stellar remnants remains crucial for unraveling the mysteries of the cosmos. Insights gained from their study have broad implications for our understanding of astrophysics, gravitational physics, and the evolution of galaxies, contributing to a deeper comprehension of the universe's structure and dynamics.
\section*{Acknowledgements}
\begin{acknowledgements}
% KMT
% Collaboration acknowledgement
E.O.O. is grateful for the support of
grants from the
Willner Family Leadership Institute,
André Deloro Institute,
Paul and Tina Gardner,
The Norman E Alexander Family M Foundation ULTRASAT Data Center Fund,
Israel Science Foundation,
Israeli Ministry of Science,
Minerva,
NSF-BSF,
Israel Council for Higher Education (VATAT),
Sagol Weizmann-MIT,
Yeda-Sela,
and the Rosa and Emilio Segre Research Award.
This research was supported by the Institute for Environmental Sustainability (IES) and The André Deloro Institute for Space and Optics Research at the Weizmann Institute of Science.

This research has made use of the KMTNet system
operated by the Korea Astronomy and Space Science Institute
(KASI) at three host sites of CTIO in Chile, SAAO in South
Africa, and SSO in Australia. Data transfer from the host site to
KASI was supported by the Korea Research Environment
Open NETwork (KREONET). This research was supported by KASI
under the R\&D program (project No. 2024-1-832-01) supervised
by the Ministry of Science and ICT.
% Individual acknowledgements
W.Zang, H.Y., S.M., R.K., J.Z., and W.Zhu acknowledge support by the National Natural Science Foundation of China (Grant No. 12133005). 
W.Zang acknowledges the support from the Harvard-Smithsonian Center for Astrophysics through the CfA Fellowship. 
J.C.Y. and I.-G.S. acknowledge support from U.S. NSF Grant No. AST-2108414. 
Y.S. acknowledges support from BSF Grant No. 2020740.
Work by C.H. was supported by the grants of National Research Foundation of Korea (2019R1A2C2085965 and 2020R1A4A2002885). 

%MOA
The MOA project is supported by JSPS KAKENHI Grant Number JP24253004, JP26247023,JP16H06287 and JP22H00153.
\end{acknowledgements}

\newpage
\bibliography{papers2}{}
\bibliographystyle{aasjournal}

%% This command is needed to show the entire author+inst list when


\end{document}